\documentclass{pasj00}
\draft

\begin{document}
\SetRunningHead{M. Sugizaki et al.}{Suzaku Image Deconvolution}
\Received{2007/xx/xx}
\Accepted{2008/xx/xx}

\title{Deconvolution of Images Taken with the Suzaku X-ray Imaging
Spectrometer}

\author{
  Mutsumi \textsc{Sugizaki}\altaffilmark{1,2}
  Tuneyoshi \textsc{Kamae}\altaffilmark{1}
  and
  Yoshitomo \textsc{Maeda}\altaffilmark{3}}
\altaffiltext{1}{
  Stanford Linear Accelerator Center, 2575 Sand Hill Road, California 94025, USA}
\email{kamae@slac.stanford.edu}

\altaffiltext{2}{
  Cosmic Radiation Laboratory, RIKEN, 
  2-1 Hirosawa, Wako, Saitama 351-0198}
\email{sugizaki@crab.riken.jp}

\altaffiltext{3}{
  Institute for Space and Astronautical Science, 
  Japan Aerospace Exploration Agency, Sagamihara, Kanagawa 229-8510
}
\email{ymaeda@astro.isas.jaxa.jp}


\maketitle

\begin{abstract}

We present a non-iterative method to deconvolve the spatial response
function or the point spread function (PSF) from images taken with the
Suzaku X-ray Imaging Spectrometer (XIS).  The method is optimized for
analyses of extended sources with high photon statistics.
Suzaku has four XIS detectors each with its own X-ray CCD and
X-Ray Telescope (XRT) and has been providing unique opportunities in 
spatially-resolved spectroscopic analyses of extended objects. The 
detectors, however, suffer from broad and position-dependent PSFs 
with their typical half-power density (HPD) of about $110''$. 
In the authors' view, this shortcoming has been preventing 
the high collecting area and high spectral resolution of Suzaku 
to be fully exploited. 
The present method is intended to recover spatial resolution to $\sim 15''$
over a dynamic range around 1:100 in the brightness without assuming any 
source model. Our deconvolution proceeds in two steps:  An XIS image is 
multiplied with the inverse response matrix calculated from its 
PSF after rebinning CCD pixels to larger-size tiles 
(typically $6''\times 6''$); 
The inverted image is then adaptively smoothed to obtain the final
deconvolved image.
The PSF is modeled on a ray-tracing program and an observed
point-source image.  
The deconvolution method has been applied to images of Centaurus A,
PSR B1509-58 and RCW 89 taken by one XIS (XIS-1). The results 
have been compared with images obtained with Chandra to conclude that 
the spatial resolution has been recovered to $\sim 20''$
down to regions where surface brightness is about 1:50 of the 
brightest tile in the image. 
We believe the spatial resolution and the dynamic range can be improved 
in the future with higher fidelity PSF modeling and higher precision 
pointing information. 

\end{abstract}

\section{Introduction}
\label{sec:intro}

Three X-ray astronomical observatories, Chandra, XMM-Newton, and 
Suzaku, are operational in orbit now. 
All have one or more X-ray imaging spectrometers, 
each made of an X-ray mirror and an X-ray CCD array. 
Among the three, Suzaku has the second largest effective area at
higher energies (590 cm$^2$ at 8 keV) when its four X-ray imaging
spectrometers (XISs) are combined, the best energy resolution (FWHM
$\sim 130$ eV at 5.9 keV) and the lowest background rate ($\sim
1.0\times 10^{-7}$ ${\rm counts~ s^{-1}~ keV^{-1}~ arcmin^{-2}~
cm^{-2}}$ at 6 keV) \citep{Mitsuda2007}.
However, these advantages are often compromised by its relatively poor
spatial resolution.  The half power diameter (HPD) of the Chandra,
XMM-Newton and Suzaku X-ray imaging spectrometers are about $1''$, $8''$, and $110''$,
respectively.  The point spread function (PSF) of the four Suzaku XISs 
is not only large and complex but depends on the position 
in the focal plane making analyses of source-rich regions or extended 
sources difficult \citep{Serlemitsos2007}. 
Whereas the HPD of the Suzaku X-ray telescopes (XRTs) is large, their 
PSF has a sharp core with an exponential radial 
profile of characteristic spread $\sim 10''$ \citep{Serlemitsos2007}.  
The pixel size of the XIS CCDs is $1''.04$ and much smaller than 
the width of the PSF \citep{Koyama2007a}. If a high number of photons 
are available, spatial resolution of XIS can be recovered 
to the PSF core size ($\simeq 10''$). We present here a first attempt 
to improve spatial resolution of Suzaku XIS by deconvolving 
the PSF.  

In astronomy, many attempts have been made in the past to obtain
better spatial information from observed data as reviewed by
\citet{Starck2002} and \citet{Puetter2005}. 
Several attempts have been reported for ASCA X-ray images 
which had a PSF similar to that of Suzaku.
\citet{White2000} employed a maximum-likelihood method to deconvolve
images of the Gas Imaging Spectrometer (GIS) aboard ASCA.  They
reconstructed spatially resolved spectrum by first
deconvolving energy-selected images with a maximum-likelihood 
method and then reassigning individual observed photons to a position 
in the deconvolved image space with a Monte-Carlo method.
\citet{Hwang1997} applied a Richardson-Lucy method to Solid-state
Imaging Spectrometer (SIS) images of a supernova remnant in 
selected energy bands.  In these methods, images are deconvolved by 
assuming a model of extended emission or 
a collection of point sources.

Because of the high through-put provided by the four Suzaku XISs,
photon statistics does not limit fidelity of image deconvolution for
bright targets.  The inverse matrix method used in this work 
can recover the true image 
if the response matrix is known accurately and if the photon statistics 
is high in the region of interest.  The method has a merit that
it does not require any prior model to fit and hence relatively free 
of systematic bias. It is best suited for complex extended sources. 

Our prime targets are galaxy clusters, pulsar wind nebulae and 
supernova remnants. The response function of the detector 
must be modeled accurately and the noise due to Poisson fluctuation 
must be controlled well to reproduce the image faithfully in 
typical brightness variation (an order of magnitude) 
in these extended objects. 
Since the PSF of Suzaku XIS is complex, the 
method requires intense labor. Once a procedure is 
established, however, this method can be set up for automatic 
deconvolution with minimum human intervention. 

The goal of this deconvolution is to recover spatial resolution to $\sim 15''$ 
while keeping fidelity in a dynamic range around 1:100 in the brightness. 
Deconvolution proceeds in two steps:  Each XIS image is 
multiplied with the inverse response matrix calculated from its 
PSF after rebinning the raw CCD image to larger tiles 
(typically $6''\times 6''$). This is required to secure high photon counts 
in each image elements. The inverted image is then adaptively smoothed 
to obtain the final deconvolved image.

When this work began, the released XIS images were degraded 
due to wobbling of satellite pointing \citep{Serlemitsos2007}.  
We have added one more step to correct for this 
pointing error by using a bright point source. 
We have modeled the PSF in two ways: one based on 
the ray-tracing simulation program developed by the Suzaku team 
\citep{Ishisaki2007} and the other by fitting the image of 
Centaurus A. The two PSFs agree well for XIS-1 but not for the others. 
For this reason we use only one of the 4 XISs when deconvolving 
the images of Centaurus A, PSR B1509-58 and RCW 89. 

The paper proceeds as follows.  We present our image-deconvolution
method in section \ref{sec:dec_method}. The Suzaku
data processing and the correction prodedure for the pointing error 
are described in section
\ref{sec:data}.  The PSF modeling is explained in section
\ref{sec:psfcal}.  We then apply our deconvolution method to XIS-1
images and compare the results with corresponding images 
of Chandra ACIS in
section \ref{sec:demo}.  Conclusions and our future plan are given in
section \ref{sec:conc}.

\section{Deconvolution by Inverse Response Matrix and Adaptive Smoothing}
\label{sec:dec_method}

We here denote an image derived by multiplying an observed image 
with the inverse response matrix as an ``inverted image''.  
When an ``inverted image'' is adaptively smoothed, we called it 
an ``deconvolved image.''

\subsection{Inverse Matrix Method}

The relation between an observed image $d(\vec{x})$
and a true image on the sky $s(\vec{x})$ is represented by
\begin{equation}
d(\vec{r})=\int s(\vec{r'})P(\vec{r'};\vec{r})d\vec{r'}
+n(\vec{r}) \label{equ:obs_cont}
\end{equation}
where $P(\vec{r'};\vec{r})$ is PSF at the source position $\vec{r'}$ 
and $n(\vec{r})$ represents collection of ``noises'' like 
Poisson fluctuation in photon counts, instrumental noise, 
and imperfect PSF modeling.

An observed image taken by a pixelized imager such as a CCD is
represented by a finite dimension vector $\vec{d}=\{d_i\}_{i=1...N}$,
where $N$ is the number of pixels used in the inversion. 
We combined multiple CCD pixels into a tile to control 
Poisson fluctuation. The image region contains $64\times 64$ tiles 
and hence $N=4096$.

The discretized version of equation (\ref{equ:obs_cont}) is
\begin{equation}
d_i = \sum_j p_{ij} s_j + n_i ~~~~~(i=1...N), \label{equ:obsdist}
\end{equation}
where $\vec{p}_j = \{p_{ij}\}_{i=1...N}$ is the PSF at
the $j$-th tile, $\vec{r}_j$. The response matrix, 
is then a matrix consisting of N PSF-vectors and 
equation (\ref{equ:obsdist}) becomes 
\begin{equation}
\vec{d} = P \vec{s}+\vec{n},
\end{equation}
where $P$ is,
\begin{equation}
P= \{\vec{p}_1 \vec{p}_2 ... \vec{p}_N \} \label{equ:rspmat}
\end{equation}
The inverse response matrix is represented by $P^{-1}$. 
The inverted image $\vec{s'}=\{s'_i\}_{i=1...N}$ is 
calculated by multiplying the inverse response matrix $P^{-1}$
(dimension $N\times N$)
to a raw image vector $\vec{d}$ as, 
\begin{equation}
\vec{s'}=P^{-1}\vec{d} = P^{-1}(P\vec{s}+\vec{n}) = \vec{s}+P^{-1}\vec{n}
\end{equation}

When the noise term $P^{-1} \vec{n}$ is negligible relative to the signal
$\vec{s}$, the inverted image is a good approximation to the true
image.  However, this is not true generally.  
In many cases, the noise term, $P^{-1} \vec{n}$, dominates over the
signal in the inverted image as described in \cite{Starck2002, Puetter2005}. 
We control this noise term by adaptive smoothing as will be described in the 
next subsection.

\subsection{Adaptive Smoothing of Inverted Image}

We employ a technique known as adaptive smoothing to control 
the noise term in the inverted image while keeping highest 
spatial resolution compatible with photon statistics.  Adaptive
smoothing is a generic smoothing method where the spatial resolution 
is balanced to the signal-to-noise ratio expected for each 
position in the image (e.g. \cite{Lorenz1993,Huang1996,2006MNRAS.368...65E}).  
It can be optimized to a strategy: we take a uniform significance approach 
in which the smoothing scale is adjusted so that the smoothed data have 
a similar signal-to-noise ratio everywhere in the image. 
This approach has been employed in AKIS \citep{Huang1996} and ASMOOTH
\citep{2006MNRAS.368...65E}. In the present adaptation, we smooth 
the inverted image not the observed image. 

Our adaptive smoothing is performed by multiplying a multi-scale
smoothing kernel, ${\cal K}(\sigma, \vec{r})$ tile-by-tile over 
the inverted image plane.
The relation between 
the input image $s'(\vec{r})$ and the smoothed image $s^\star(\vec{r})$ 
is represented by
\begin{equation}
\vec{s^\star}(\vec{r}) 
= \int s'(\vec{r'}) {\cal K}(\sigma(\vec{r});\vec{r'}-\vec{r}) d\vec{r'}
\end{equation}
We here note that the width $\sigma(\vec{r})$ 
is defined in the smoothed image space $\vec{r}$,
not in the input image space $\vec{r'}$.

We use a Gaussian kernel.
The smoothing matrix ($N\times N$ dimension) for pixelized image is 
\begin{eqnarray}
F 
&=&  \left\{{\cal K}(\sigma(\vec{r_i}), \vec{r_j}-\vec{r_i})\right\}_{i=1...N, j=1...N}\\
&=& \left\{\frac{1}{2\pi\sigma_i^2} \exp\left(-\frac{|\vec{r_j}-\vec{r_i}|^2}{2\sigma_i^2}\right)\right\}_{i=1...N, j=1...N}
\end{eqnarray}
The smoothed image $\vec{s^\star}$ and  
the noise image $\vec{n^\star}$ are represented by
\begin{equation}
\vec{s^\star} = F P^{-1} \vec{d} = T \vec{d}
\end{equation}
\begin{equation}
\vec{n^\star} = F P^{-1} \vec{n} = T \vec{n}
\end{equation}
where $F P^{-1} = T = \{t_{ij}\}_{i=1...N,j=1...N}$.
The signal-to-noise ratio at each pixel $\{\it{SNR}_i\}_{i=1...N}$
is then,
\begin{equation}
{\it SNR_i} 
= \frac{s^\star_i}{n^\star_i}
= \frac{T\vec{d}}{T\vec{n}}=\frac{\sum_j t_{ij}d_j}{\sum_j t_{ij}n_j}
\label{equ:snr_def}
\end{equation}
When Poisson noise is dominant and $d_i$ is large, 
$n_i = \sqrt{d_i}$.
Since the Poisson noise of photon counts at each pixel can be regarded 
as mutually independent, 
the equation (\ref{equ:snr_def}) is reduced to the following equation.
\begin{equation}
{\it SNR}_i 
=\frac{\sum_{j} t_{ij} d_j}{\sqrt{\sum_{j} (t_{ij} n_j)^2}}
=\frac{\sum_{j} t_{ij} d_j}{\sqrt{\sum_{j} t_{ij}^2 d_j }} \label{equ:snr}
\end{equation}

\subsection{Search for Optimal Smoothing Scales}
\label{sec:aks_search}

Our strategy is to find a set of smoothing widths $\{\sigma_i\}_{i=1...N}$
so that the signal-to-noise ratio
$\{{\it SNR}_i\}_{i=1..N}$ agree
with a given preset value ${\it SNR}_{\rm opt}$ at every pixel on the image. 
We developed a deconvolution program to search for an optimum 
iteratively.
A similar method has been used in ASMOOTH \citep{2006MNRAS.368...65E}.  
The program starts to test smoothing widths $\{\sigma_i\}_{i=1...N}$ from an
initial minimum value.  The larger the smoothing width is, 
the lower the spatial resolution of the inverted image will be.
The program increases each of
$\{\sigma_i\}_{i=1...N}$ step by step, then stops to increase
$\sigma_i$ when ${\it SNR}_i$ reaches the preset desired value.  This
step is repeated until ${\it SNR}_i$ reaches the desired value for all
pixels or $\sigma_i$ reaches a preset maximum value.

The deconvolution program takes four parameters: the desired
signal-to-noise ratio ${\it SNR}_{\rm opt}$, 
the minimum smoothing scale $\sigma_{\rm min}$, 
the maximum smoothing scale $\sigma_{\rm max}$, 
and the increment for each iterative step $\Delta\sigma$.  
Among the four parameters, the deconvolution result is 
sensitive only to the signal-to-noise ratio ${\it
  SNR}_{\rm opt}$.  The
minimum scale $\sigma_{\rm min}$ and 
the increment for each step $\Delta\sigma$
are chosen to be smaller than 1 tile size of the input image 
($=6''$) to achieve the maximum resolution.
The maximum scale $\sigma_{\rm max}$ is chosen as large as the 
standard deviation of the PSF ($\simeq 30''$).  
We use the parameter values, 
${\it SNR}_{\rm opt}=4.0$, 
$\sigma_{\rm min} = 2''.4$, 
$\sigma_{\rm max} = 24''$, 
$\Delta\sigma    = 1''.4$ 
in the tests of the deconvolution method in
section \ref{sec:demo}.

\section{XIS Data}
\label{sec:data}

We used Suzaku archival data released via the pipeline processing
version 1.2.2.3.
The data has been processed using the HEADAS
software version 6.1.2 released from the NASA/GSFC Guest Observer
Facility.  We extracted raw XIS images from the XIS event data
screened by the standard screening procedure.

\subsection{Correction for XRT Alignment Error}
\label{sec:attcor}

The pointing direction of the Suzaku's XRTs has been found to wobble
by $\sim 40''$ roughly synchronized with the 96-minute orbital motion
of the spacecraft \citep{Serlemitsos2007}.  This wobbling 
is now understood as due to
thermal distortion.  The distortion is introduced when the side panel
\#7 on which the start trackers and the gyroscopes are mounted, is
illuminated by sun-lit Earth.  \citep{Serlemitsos2007}.  Software to
correct for the alignment error
is now available \citep{Uchiyama2008}.
However, it was not available at the time of this work and we
corrected for this error using X-ray point sources in the data.  

\begin{figure}
\begin{center}
\FigureFile(8cm,){fig1.ps}
\caption{ 
Time variation of the position of Cen A in the XIS-0 image during observation 
started at  2005-08-19 05:55:00 (UTC). The CCD pixel width is $1''.04$. 
}
\label{fig:attdist}
%

\FigureFile(5.5cm,){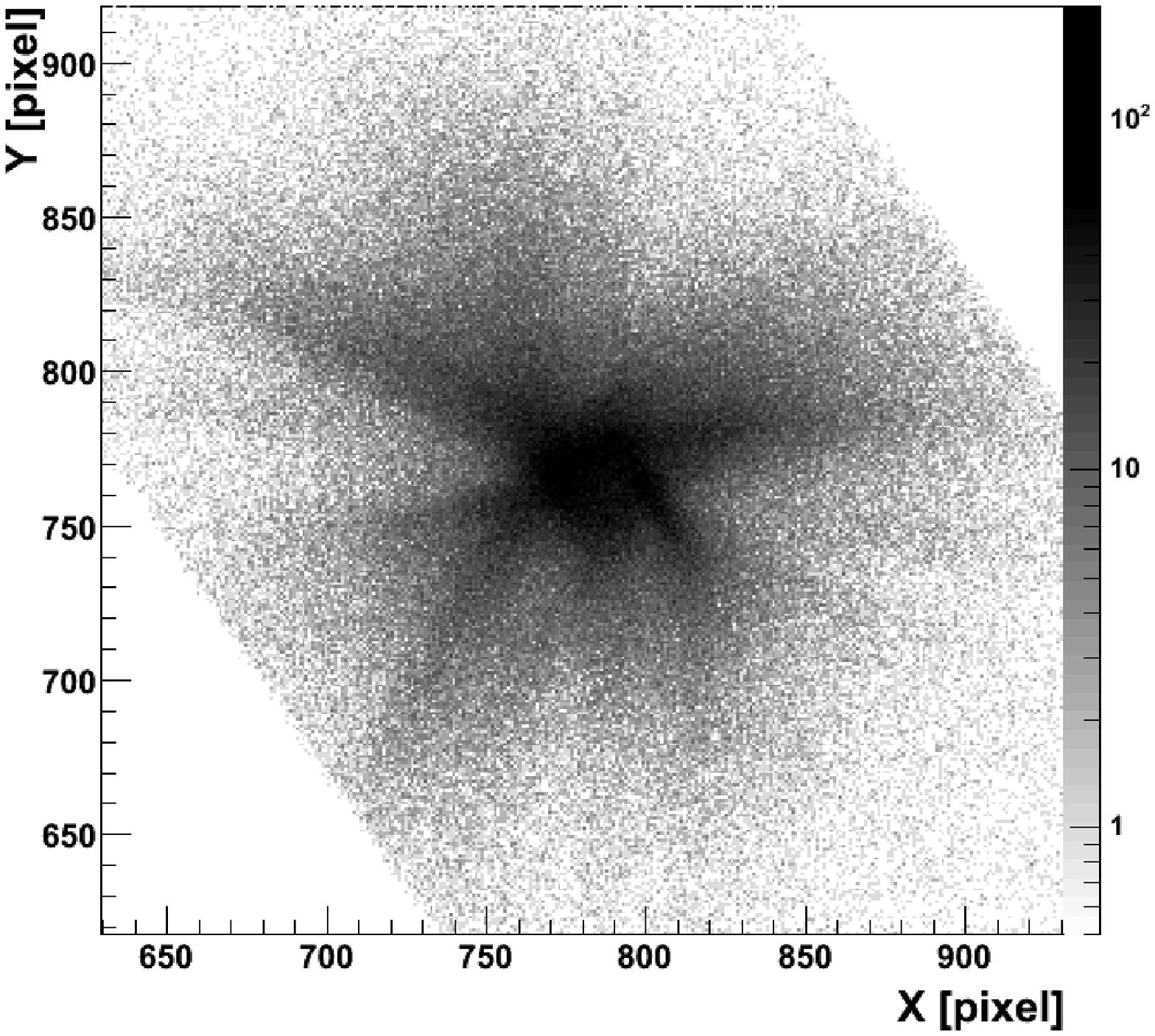}
\FigureFile(5.5cm,){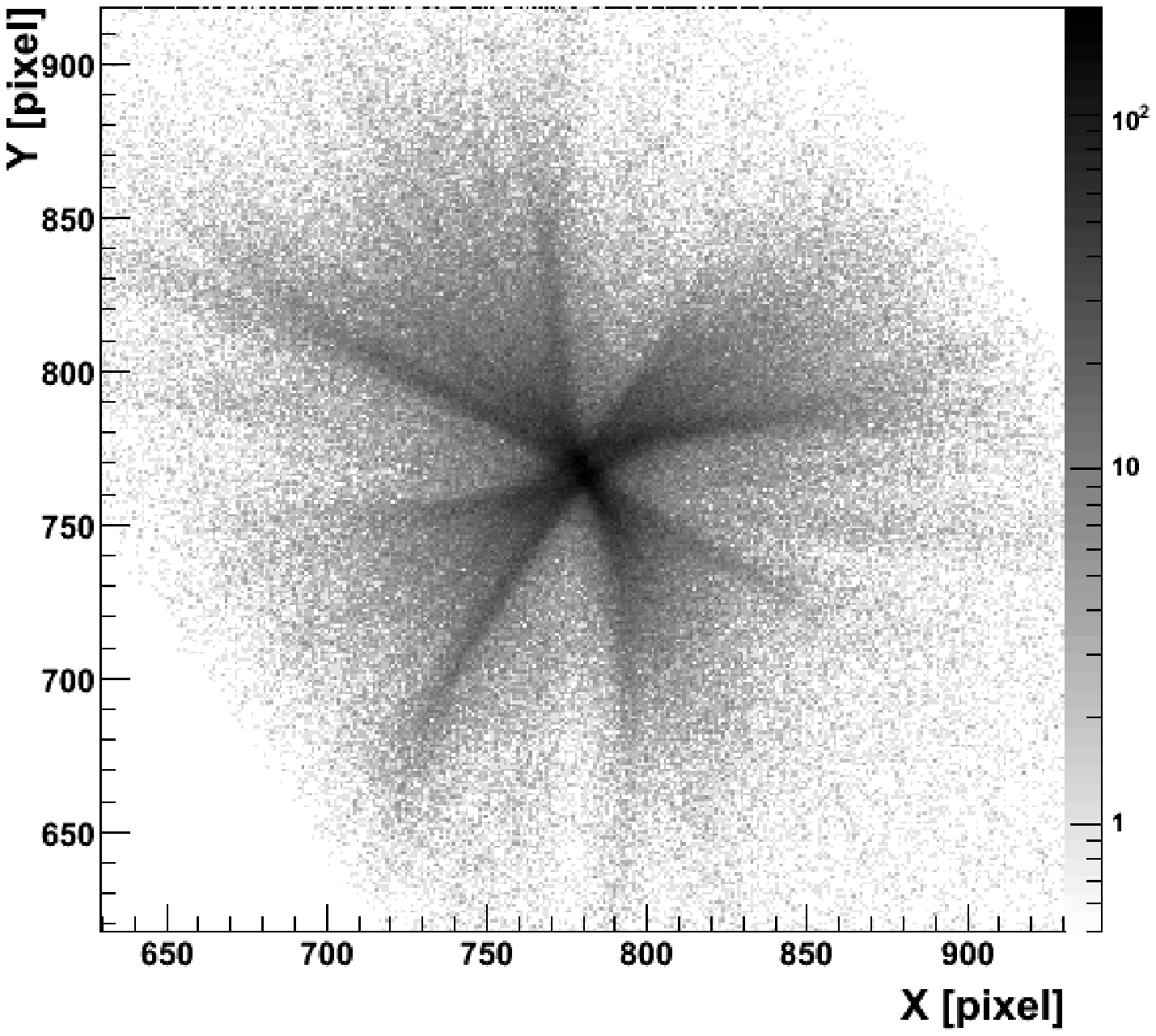}

\caption{ 
XIS-0 images of Cen A in 0.5--10 keV band 
before the XRT alignment-error correction ({\it left})
and after the correction ({\it right}).
}
\label{fig:attcorimg}
\end{center}
\end{figure}

The corrections are done in two steps.  Firstly, data was divided into
500-second intervals and the position of a bright point source is
monitored for each interval.  Figure \ref{fig:attdist} shows the time
variation of the source position in the observed Cen A data 
in the initial performance verification phase.  The periodicity
of the 96-minute period is seen clearly.  The images for 500-second 
intervals were shifted so that the Cen A align to one position.
Figure \ref{fig:attcorimg} shows the XIS-0 image of Cen A
before and after the pointing-error correction.  
The image is sharpened by the pointing-error correction.

\section{Point Spread Function Modeling}
\label{sec:psfcal}

Four XRTs have different complex position-dependent PSFs 
and they were calibrated on ground \citep{AItoh2004, KItoh2004, Misaki2004}.  
The PSFs observed in orbit are different from those measured on the ground
because of absence of gravity in orbit.  Hence, the PSF models made on
the ground calibration have to be modified for the present
application.

We constructed two PSF models: one based on the ray-tracing simulator, 
{\it{xissim}},
which we call the xissim PSF and the other on the observed Cen A image,
the observed PSF.  The xissim PSF implements the position-dependence
in the entire image area.  However xissim is based on the ground calibration 
and does not agree with the in-orbit point source images.
On the other hand, the observed PSF is available only near the 
optical axis of each XIS.
We combined the two PSF models when possible: the observed PSF 
for the central region within $6'$ from the optical axis 
and the xissim PSF for the outer region.

\subsection{Xissim PSF}
\begin{figure}
\begin{center}
\FigureFile(7.5cm,){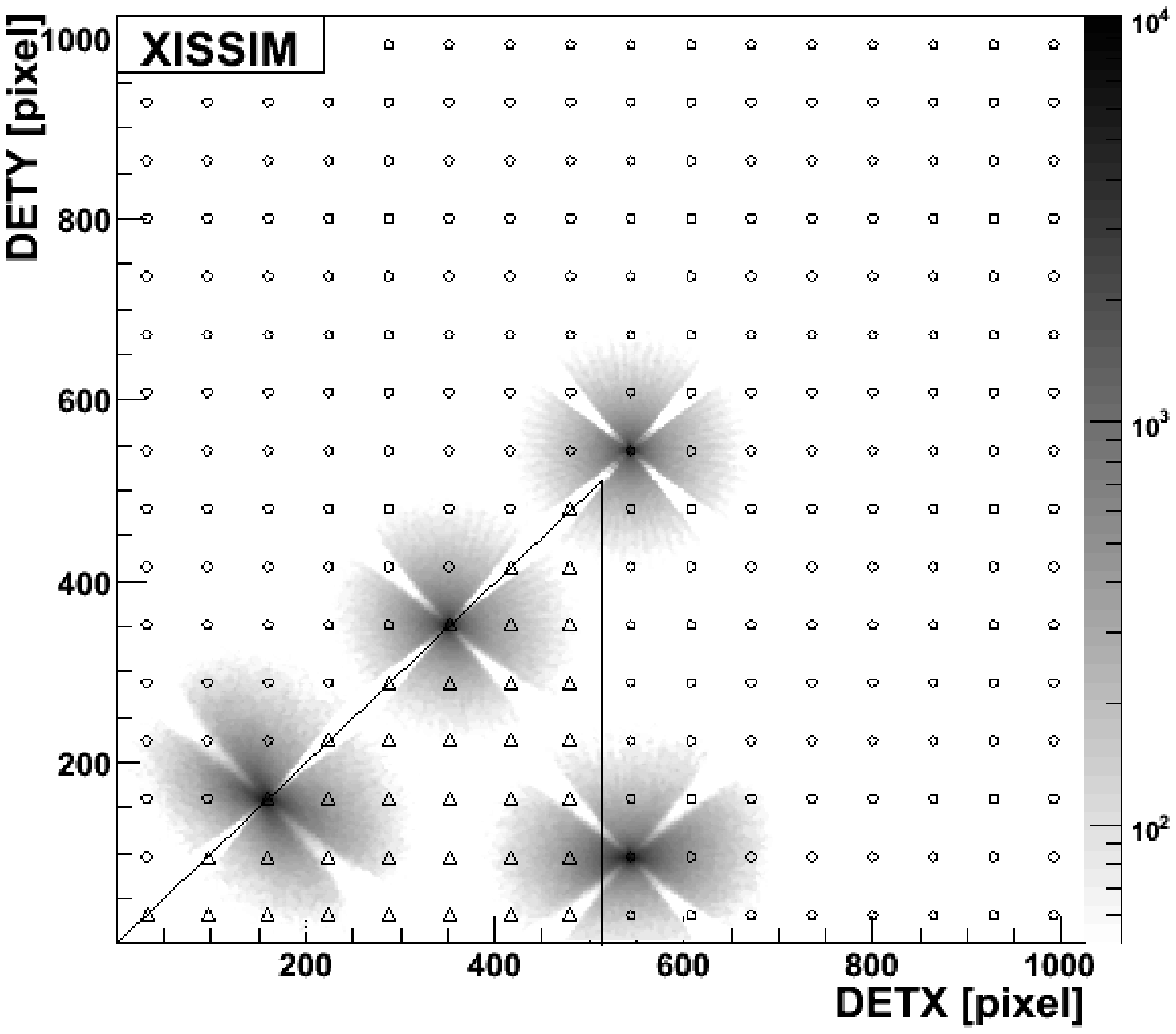}
\FigureFile(7.5cm,){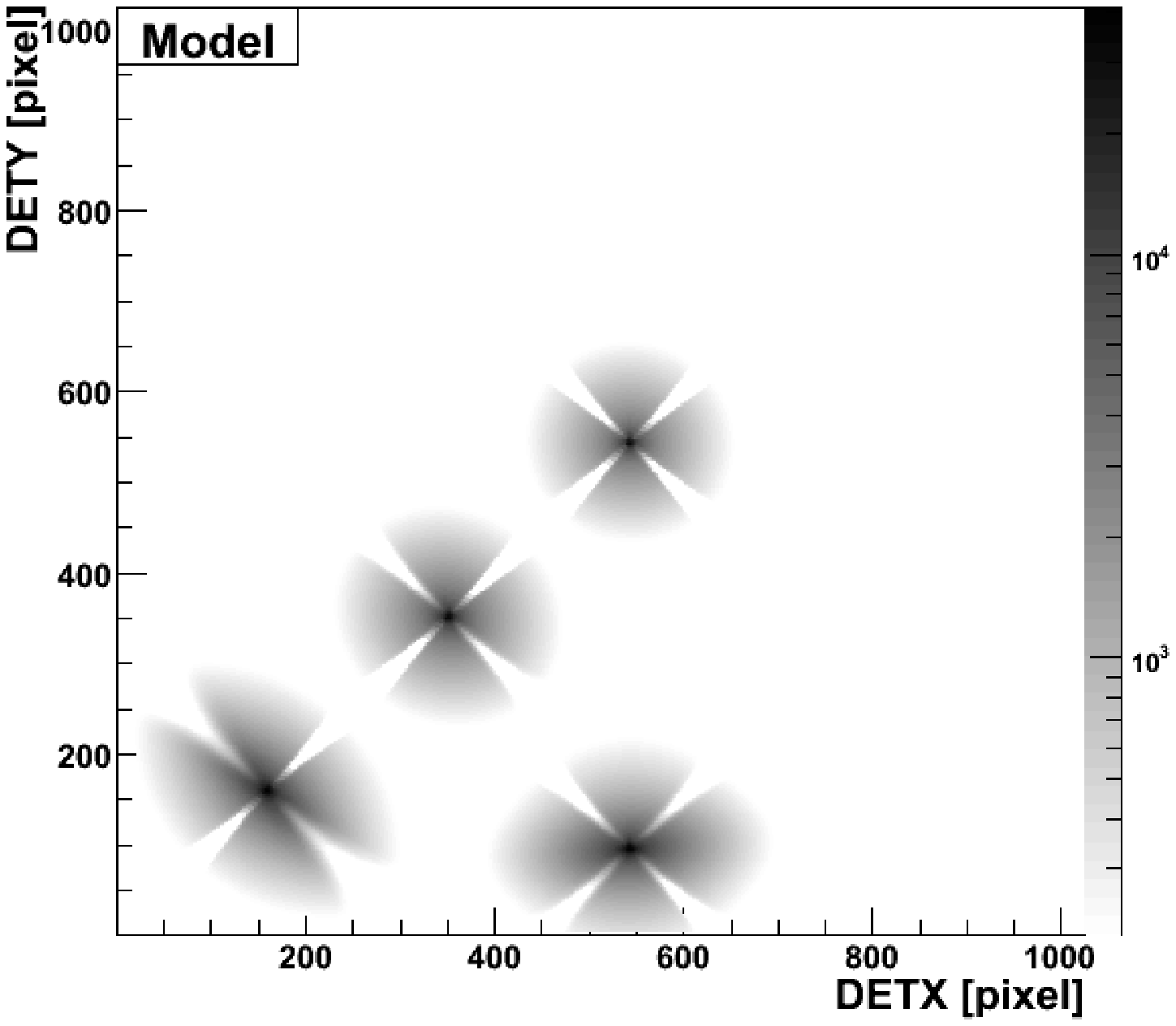}
\caption{
({\it Left}): PSF samples generated by XRT/XIS simulator, {\it xissim}. 
Open triangles are positions at which PSFs are simulated.
Open circles show the grid points where PSF can be taken from 
the symmetric point in the triangle.
({\it Right}): PSFs generated with our 
Xissim PSF model.
Gray contour colors are spaced logarithmically in the both figures.
}
\label{fig:psfsample}
\end{center}
\end{figure}

The XRT ray-tracing library,
{\it xissim}, is included in Suzaku FTOOLS in the HEADAS software
package \citep{Ishisaki2007}.  
We simulated PSF images for $16\times 16$ source locations,
\begin{equation}
({\it DETX}, {\it DETY}) = (32+64\times i, 32+64\times j)
~~~ (0\leq i < 16, ~0\leq j< 16)
\end{equation}
where each PSF represents a $1'.1\times 1'.1$ segment 
on the $17'.7\times 17'.7$ XIS image area.
We assumed that
the four quadrants of each XRT mirror are identical and
each quadrant has a mirror-symmetry relative to the median angle 
of the quadrant. 
This means that the position dependence of PSF has 
four-fold axis-symmetry around the optical axis of the XRT and 
each quadrant has internal mirror symmetry.  The position dependence 
of PSF has been simulated in the 1/8 triangular region
of the square XIS FOV.  The left panel of figure \ref{fig:psfsample}
shows the simulated position dependence of PSF.

\begin{figure}
\begin{center}
\FigureFile(9.5cm,){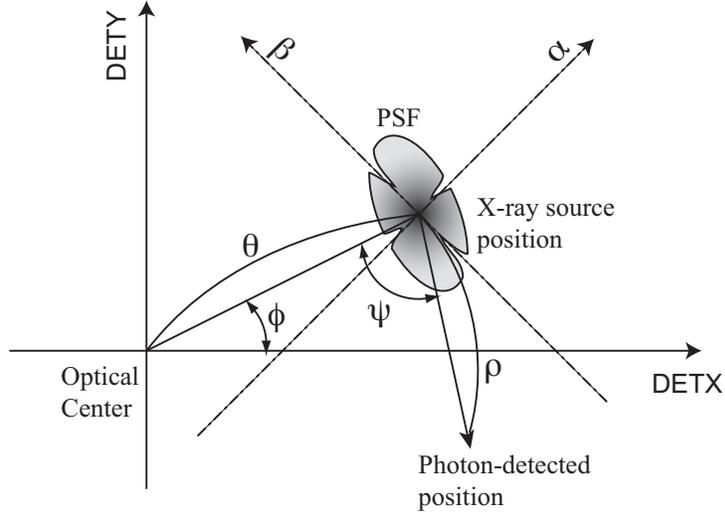}
\caption{
Coordinate definition used in PSF modeling.

}
\label{fig:psf_coord}
\end{center}
\end{figure}

The simulated PSF images have then been fitted with an analytic model
function. 
Figure \ref{fig:psf_coord} illustrates parameters used in the PSF
model: the X-ray source position $(\theta, \phi)$, the detected photon
position $(\rho, \psi)$, and coordinates for the mirror 
quadrant boundaries by $(\alpha, \beta)$
The function of the xissim PSF, ${\it PSF_{\rm xissim}}(\theta, \phi;
\rho, \psi)$, consists of a central peak profile represented by three
exponential distributions, $p(\theta, \phi; \rho, \psi)$.
The sensitivity gaps at the boundaries of the mirror quadrant
are represented by $q(\theta, \phi; \rho, \psi)$.  
They take the following formulae:
\begin{equation}
{\it PSF_{\rm xissim}}(\theta, \phi; \rho, \psi) =  p(\theta, \phi; \rho, \psi) q(\theta, \phi; \alpha, \beta)
\end{equation}
\begin{equation}
p(\theta, \phi; \rho, \psi) 
= \sum_{i=1,2,3} 
c_i \exp\left(
\frac{-\rho}{ \sqrt{a_i^2\cos^2 \psi + b_i^2\sin^2 \psi}}
\right)
\end{equation}
\begin{equation}
q(\theta, \phi; \alpha, \beta)=\left\{  
\begin{array}{ll}
\displaystyle 1 & (|\alpha|\leq k_1, ~ |\beta|\leq h_1) \\
\displaystyle \left[1+\exp\left(\displaystyle \frac{k_2(|\alpha|-k_1)-|\beta|}{k_3(|\alpha|-k_1)+k_4} \right)\right]^{-1}
& ( |\alpha|\geq|\beta|) \\
\displaystyle \left[1+\exp\left(\displaystyle \frac{h_2(|\beta|-h_1)-|\alpha|}{h_3(|\beta|-h_1)+h_4} \right)\right]^{-1}
& ( |\alpha|<|\beta|) 
\end{array}
\right.
\end{equation}
Each simulated PSF at $(\theta, \phi)$ is fitted to obtain the 17
parameters, $\{a_i, b_i, c_i\}_{i=1,2,3}$ and $\{k_i,
h_i\}_{i=1,2,3,4}$.  The right panel of figure \ref{fig:psfsample}
shows the fitted PSF.  We calculate the PSF for arbitrary position 
by interpolating from the simulated PSFs at the grid points.

\subsection{Validation of xissim PSF}
\label{sec:psfcomp}

We have validated the xissim PSF model by comparing actual
observed image of Cen A in 2--10 keV band.  
The central X-ray source is known to be point-like in 
the energy band \citep{Evans2004,Markowitz2007}.

\begin{figure}

\begin{minipage}[t]{11cm}
\FigureFile(5.5cm,){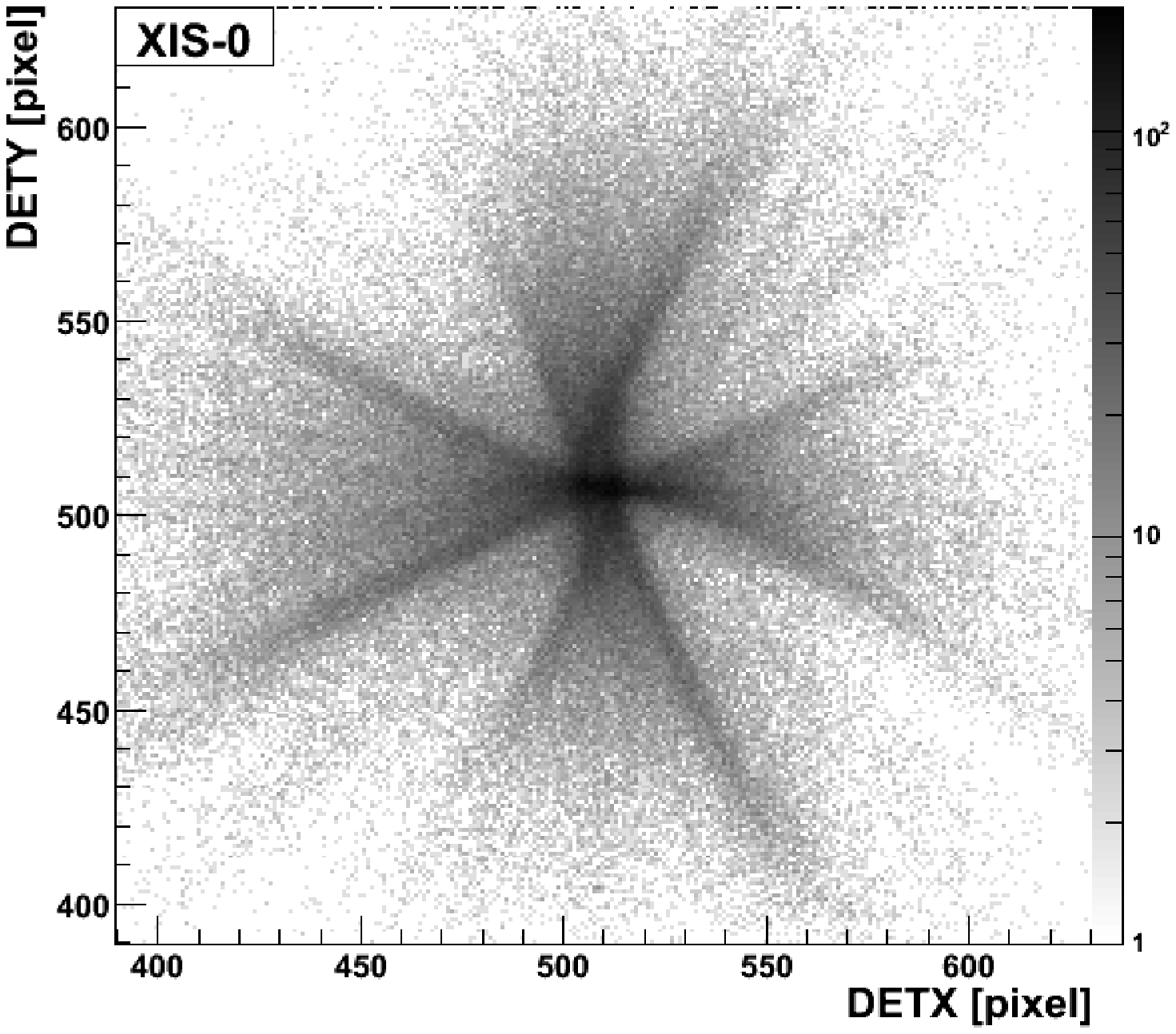}
\FigureFile(5.5cm,){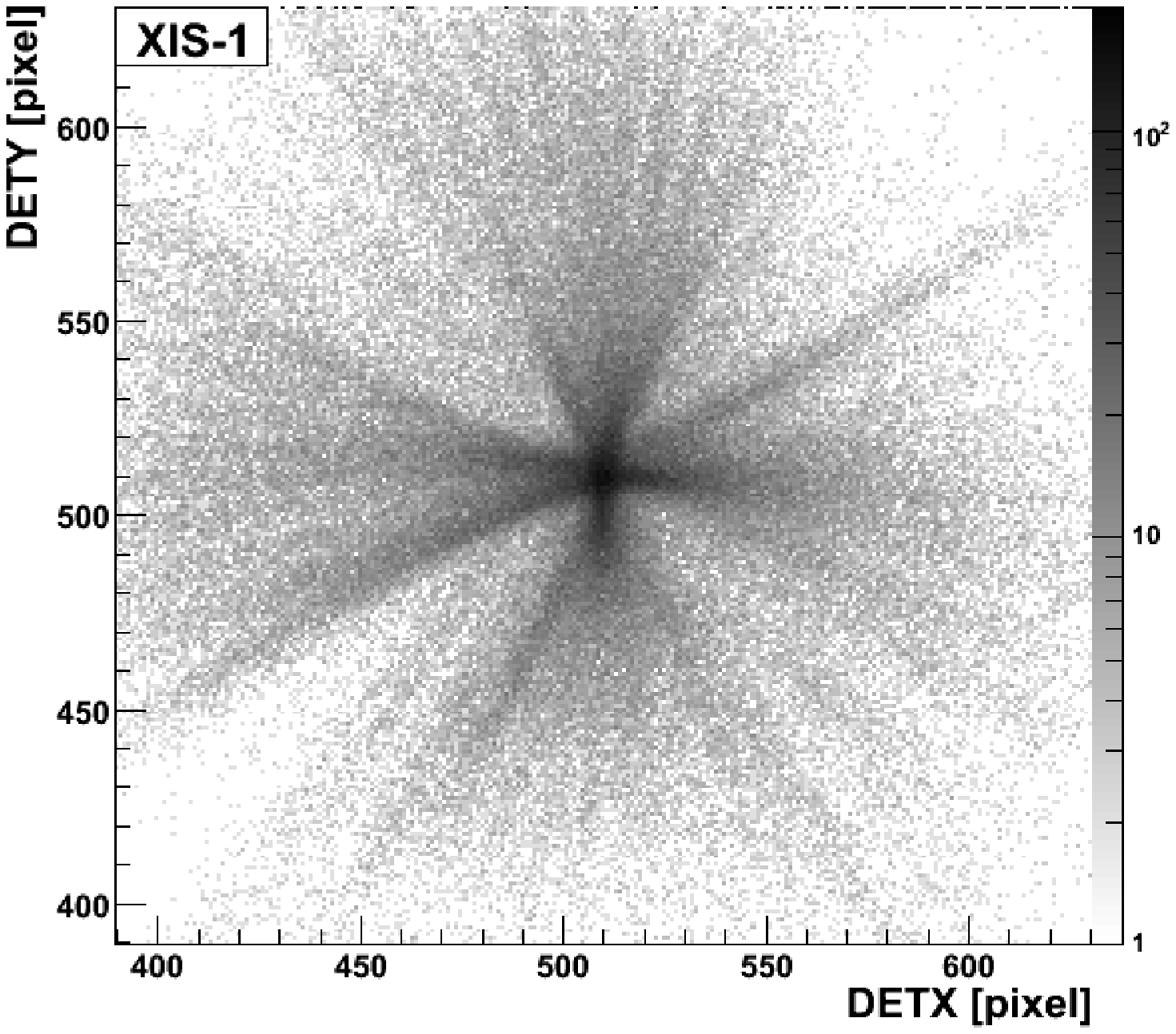}

\FigureFile(5.5cm,){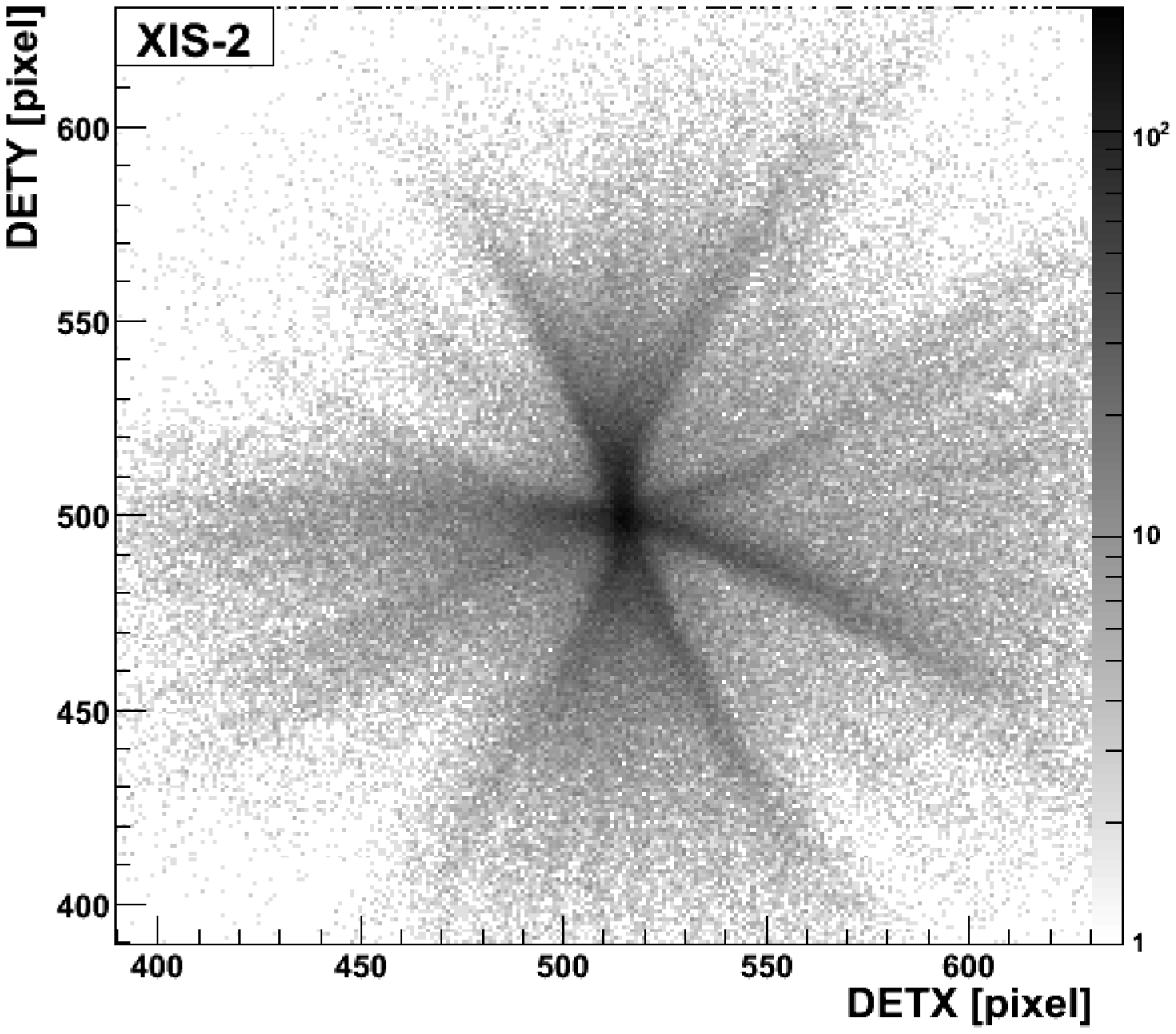}
\FigureFile(5.5cm,){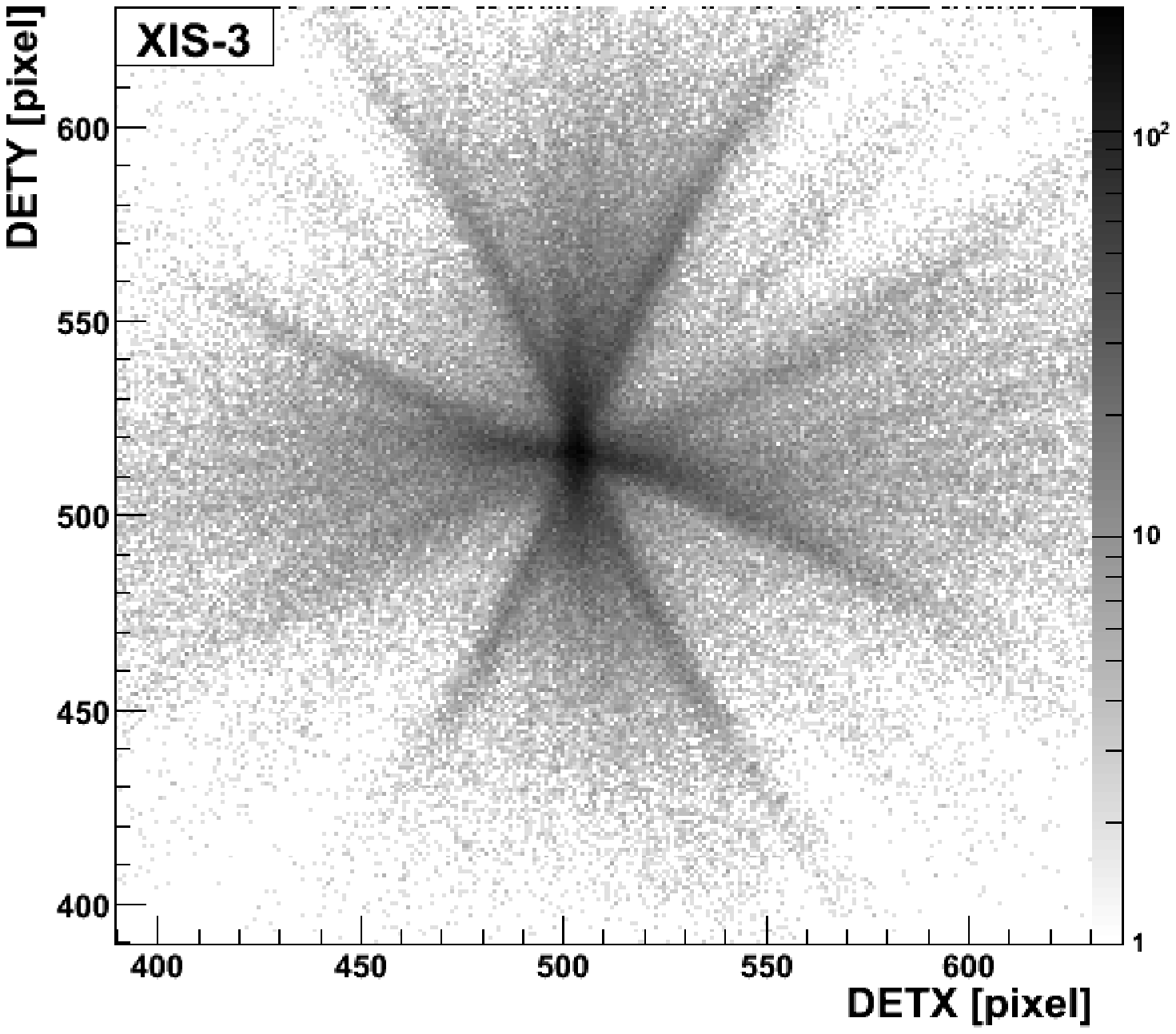}
\end{minipage}	            
\begin{minipage}[t]{5.5cm}
\FigureFile(5.5cm,){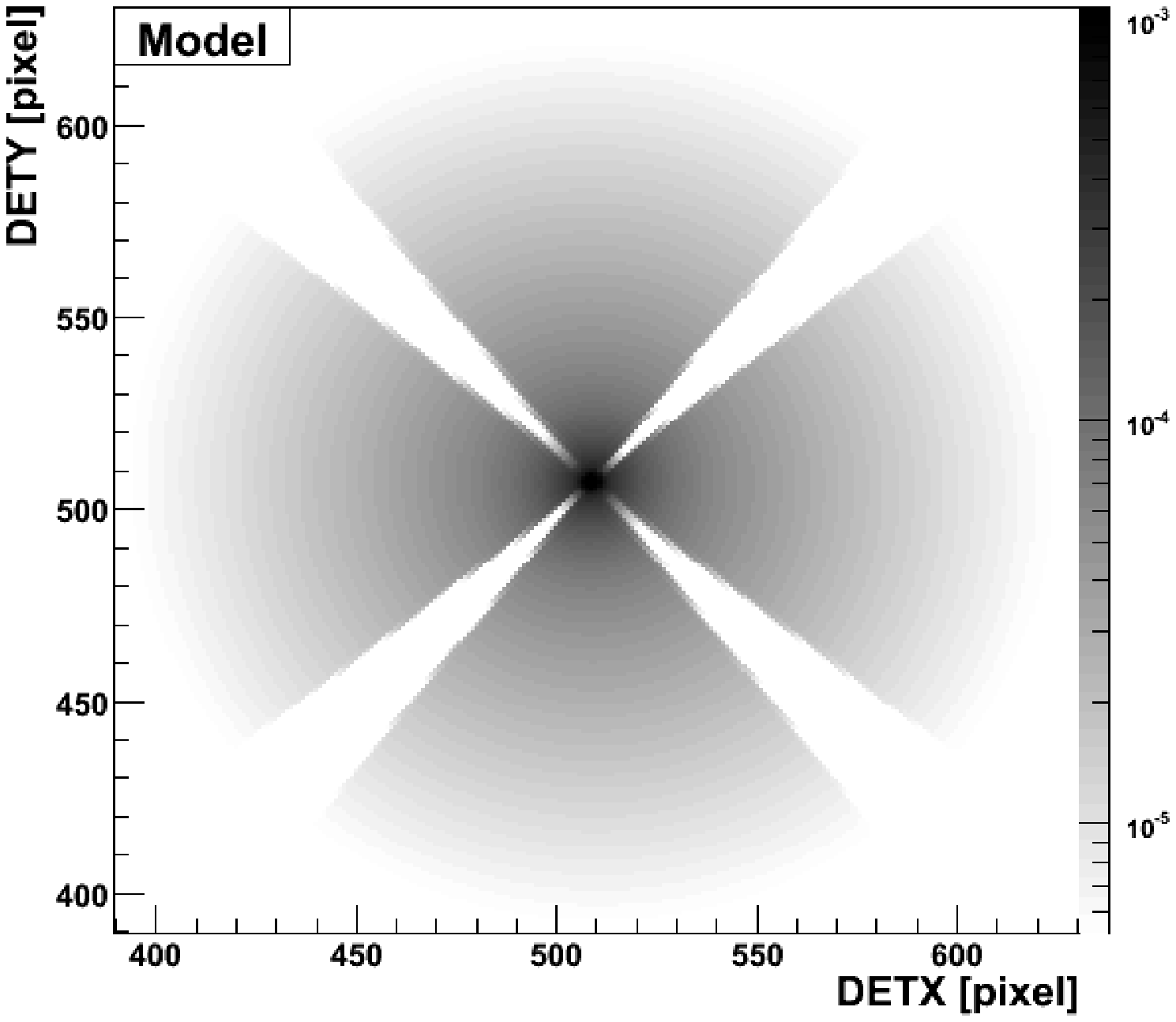}
\end{minipage}

\begin{center}

\caption{
Four XIS images of Cen A in 2--10 keV band and xissim PSF.
The gray contour colors are spaced logarithmically.
}
\label{fig:psfdatmodel}
\end{center}
\end{figure}

\begin{figure}
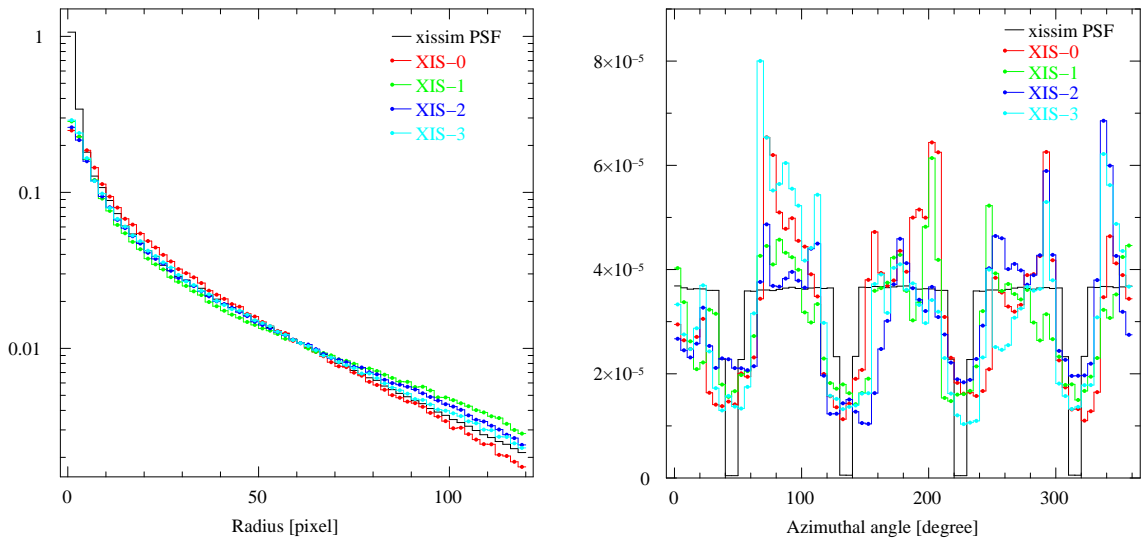

\begin{center}

\FigureFile(7.0cm,){fig6a.ps}
\hspace{0.5cm}
\FigureFile(7.35cm,){fig6b.ps}

\caption{
  ({\it Left}): Radial profiles  
  of xissim PSF and images of Cen A in 2--10 keV observed by 
  each XIS detector.
  All profiles are normalized to 0.01 at a radius of 60 pixel.  
  ({\it Right}): Azimuthal profiles 
  extracted from annulus between radii $r=40''$ and $r=60''$.  
  }
\label{fig:psf_pror}
\end{center}
\end{figure}

Figure \ref{fig:psfdatmodel} shows the Cen A images taken with the
four XISs after the XRT alignment-error correction, 
and the model PSF.
The observed Cen A image is similar to the PSF model, 
but there are apparent differences.
Figure \ref{fig:psf_pror} shows the radial profiles of the observed
Cen A images and the xissim PSF model averaged over the
$360^\circ$ azimuthal angle and the azimuthal profiles of an annulus
between radii $r=40$ pixel and $r=60$ pixel.  The azimuthally averaged
radial profile of the Cen A images agree well with the model PSF
except for the central core of radius $r<5''$.
On the other hand, the azimuthal profiles of the Cen A images are
significantly different from the model PSF, reflecting the
complicated asymmetric profiles of the real XRT-XIS system.

Among the Cen A images observed by the four XISs, that 
of the XIS-1 came closest to the model PSF.  Hence, we have decided 
to use XIS-1 to test the image deconvolution method
in section \ref{sec:demo}.

\subsection{Observed PSF}

\begin{figure}
\begin{center}

\begin{minipage}[c]{7.0cm}
\vspace{0.4cm}
\FigureFile(7.0cm,){fig7a.ps}
\end{minipage}
\hspace{0.5cm}
\begin{minipage}[c]{7.8cm}
\FigureFile(7.8cm,){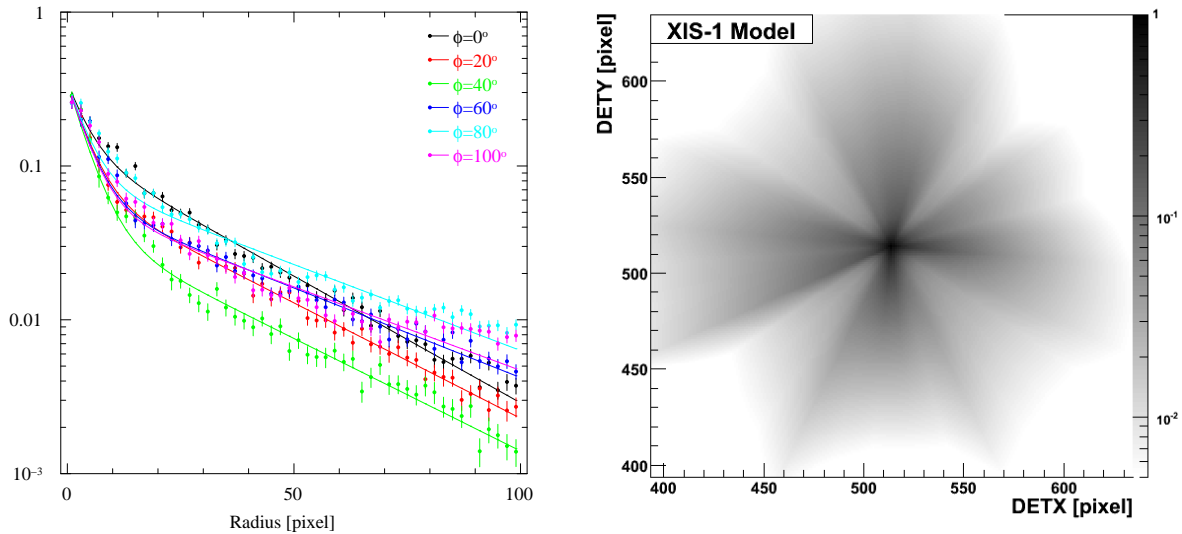}
\end{minipage}

\caption{
  ({\it Left}):
  Radial profiles of XIS-1 Cen A image in 2--10 keV band
  with statistical errors 
  and their best-fit model functions (solid lines) for 
  various azimuthal angles.
  ({\it Right}):
  Observed PSF model for central region of XIS-1 image. 
  }
\label{fig:psf_pror_xis1}
\end{center}
\end{figure}

We made another PSF model out of the observed Cen A image, 
and call it the observed PSF.
We plot radial profiles of the Cen A image for several azimuth angles in the
left panel of figure \ref{fig:psf_pror_xis1}.  These profiles are fitted
with a model functions consisting of two exponentials.  The PSF model
is formulated on the best-fit radial profiles.  The obtained
observed PSF model is shown in the right panel of figure
\ref{fig:psf_pror_xis1}.

The archived data of Cen A were taken in one pointing and hence 
position dependence of PSF could not be extracted.  Because of this, we use 
the observed PSF only for the central region of the XIS FOV in the 
deconvolution analyses described below.

\section{Application to XIS images}
\label{sec:demo}

We have applied the present deconvolution method to 
simulated XIS images as well as to observed images in the Suzaku
archival data.  We have chosen Cen A, PSR B1509-58 and RCW 89
as targets because they include both a bright point source and extended
emissions and also because there are Chandra observations 
of the same regions.  
The bright point
sources serve to correct for the XRT pointing error as has been 
described in section \ref{sec:attcor}.
The images of Chandra ACIS are used to evaluate the fidelity of the
deconvolved images.

\subsection{Deconvolution of Images}
\label{sec:procedure}

The deconvolution program is developed on a standard Linux machine in 
the C program language using
BLAS\footnote[1]{http://www.netlib.org/blas/}/
LAPACK\footnote[2]{http://www.netlib.org/lapack/} linear algebra
program libraries \citep{LAPACK} customized by ATLAS project 
\citep{ATLAS}.

The deconvolution of XIS images proceeds in the following steps.  
\begin{enumerate}
  
\item Extract a region of interest from the entire XIS image
  (1024$\times$1024 pixels covering area of $17'.8\times 17'.8$) and
  make a 64$\times$64 tiled image where each tile is combination of 
  $6\times 6$ XIS CCD pixels (each CCD pixel covers $6''.3\times 6''.3$).  
  For RCW 89, the tile size has been set larger to cover 8$\times$8 CCD pixels 
  because number of photon per tile was lower.

\item Make a response function based on the two PSF models.  
  The PSF is normalized to 1.0 so that the
  number of photons is preserved.
  We employed the observed PSF model for the region
  within $6'$ from the XRT optical axis.
  Exception is RCW 89. 
  The xissim PSF model was used for RCW 89.

\item Calculate the inverse of the response matrix.
   
\item Multiply the inverse response matrix with the tiled image

\item Smooth the response-inverted image adaptively.  The adaptive
  smoothing takes a signal-to-noise ratio,
  ${\it SNR}_{\rm opt}$ as a parameter. 
  We set ${\it SNR}_{\rm opt}=4.0$ as described in section \ref{sec:aks_search}
  for all the examples described below.

\end{enumerate}

\subsection{Test with Simulated Images}
\label{sec:simdemo}

\begin{figure}

\begin{minipage}[t]{4cm}
\FigureFile(4.0cm,){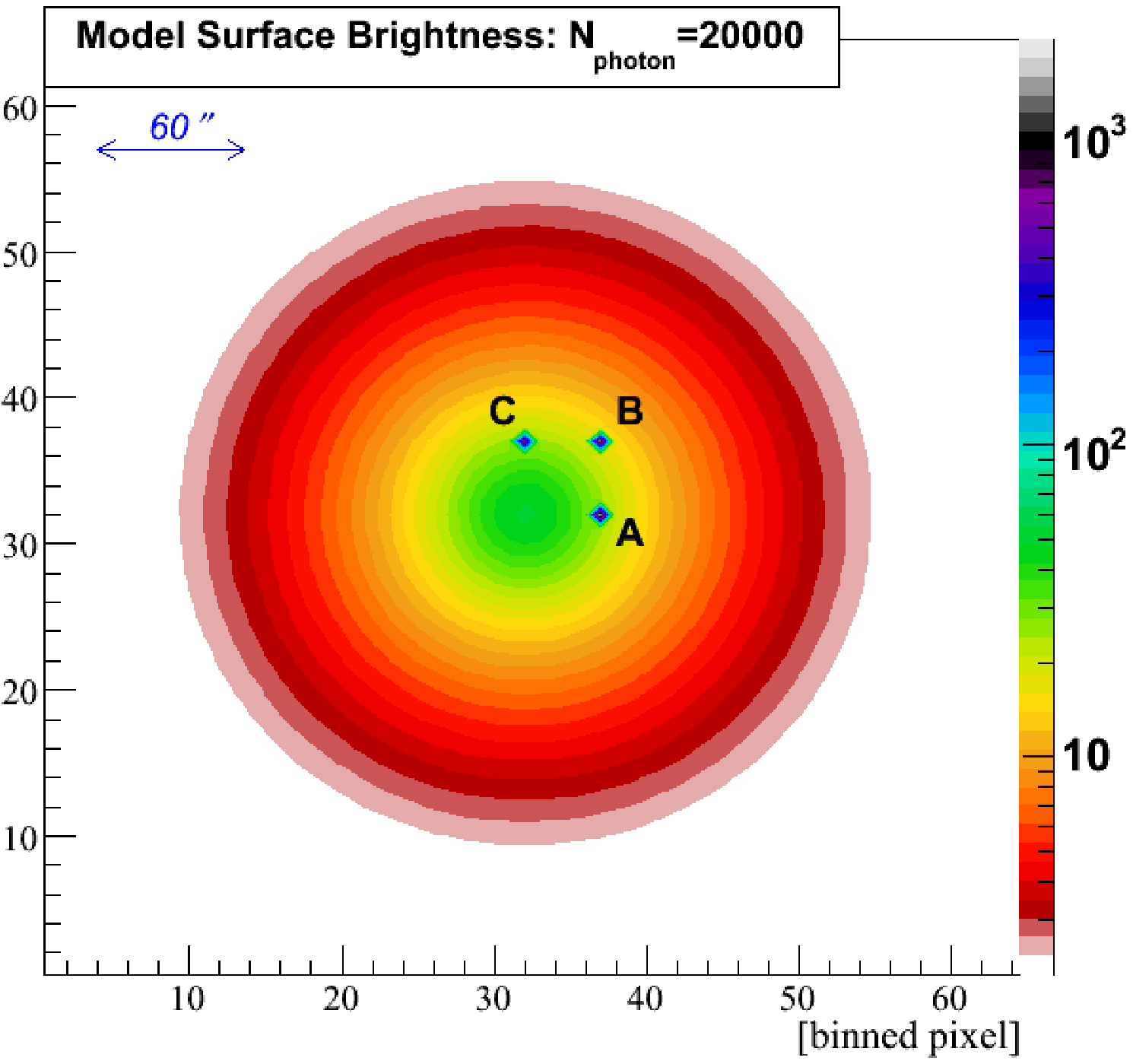}
\end{minipage}
\begin{minipage}[t]{4cm}
\FigureFile(4.0cm,){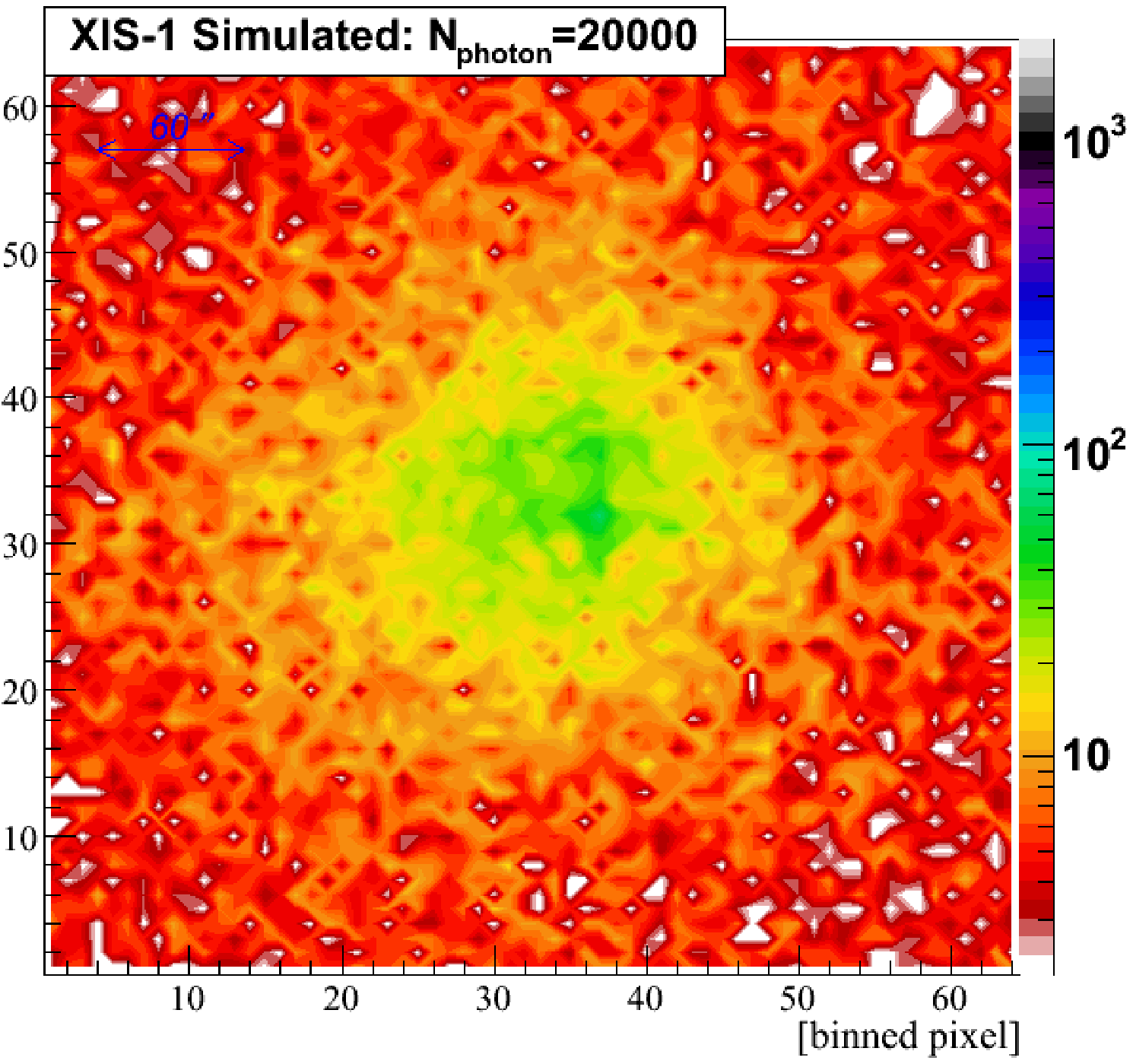}
\FigureFile(4.0cm,){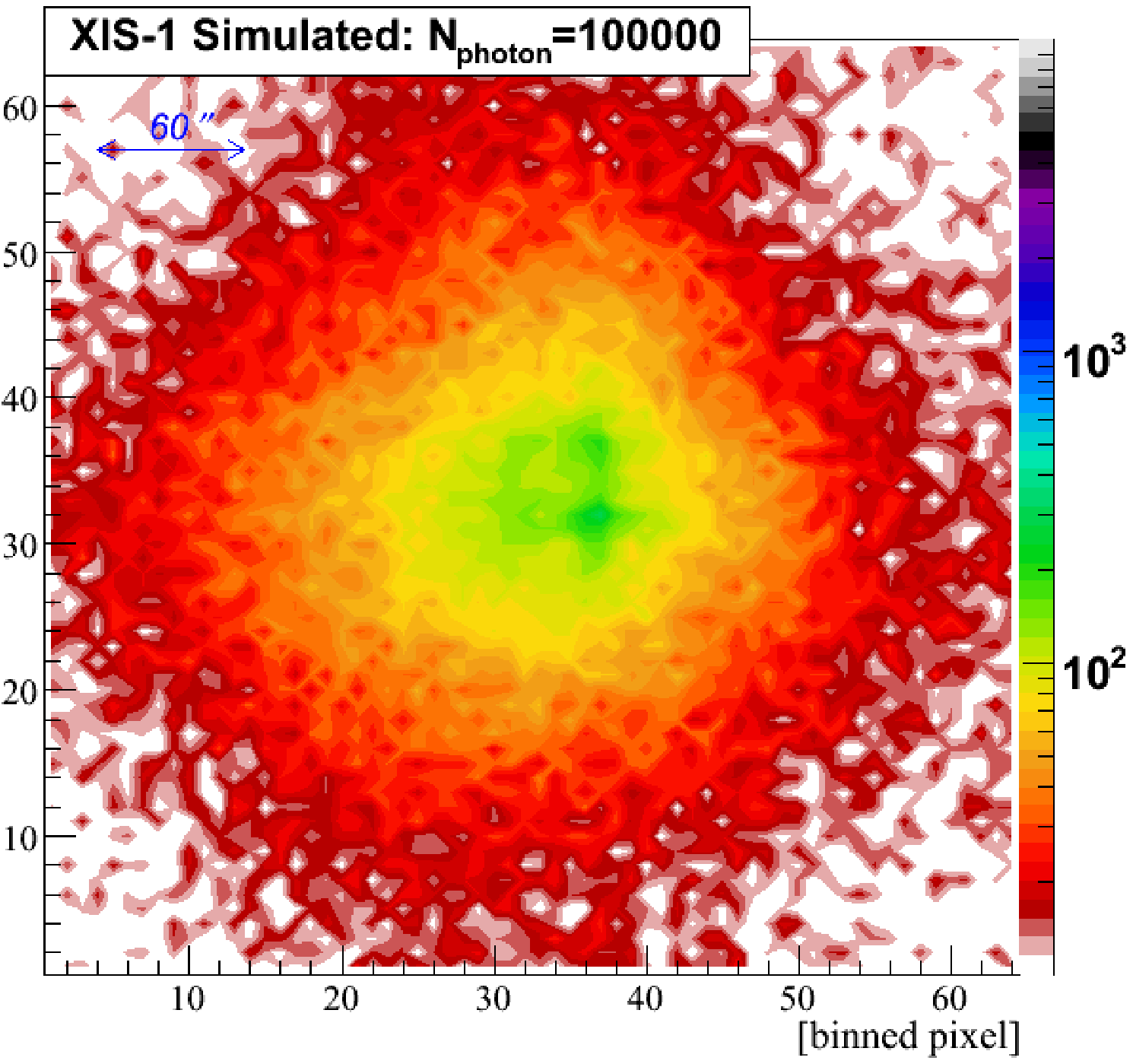}
\FigureFile(4.0cm,){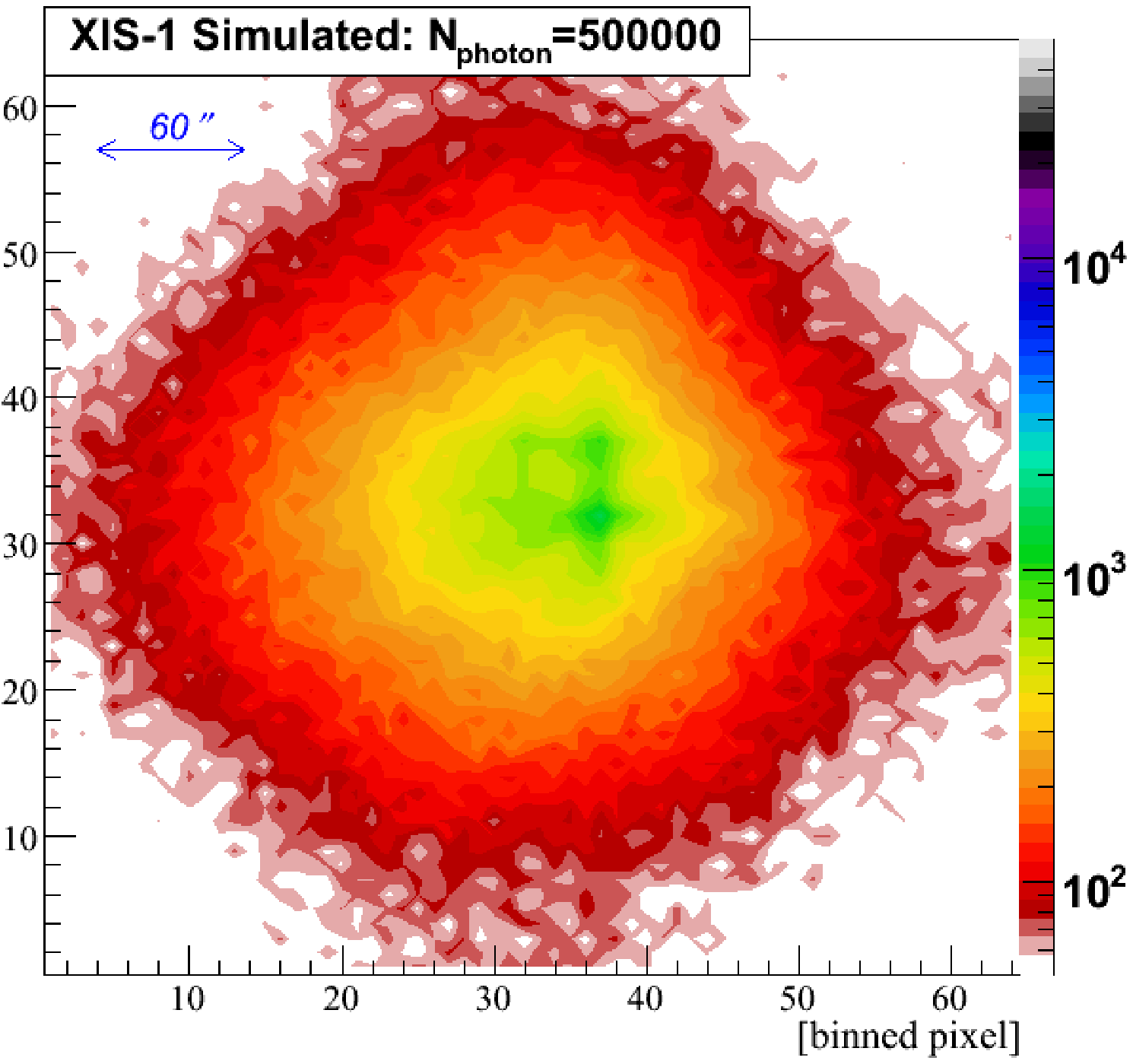}
\end{minipage}
\begin{minipage}[t]{4cm}
\FigureFile(4.0cm,){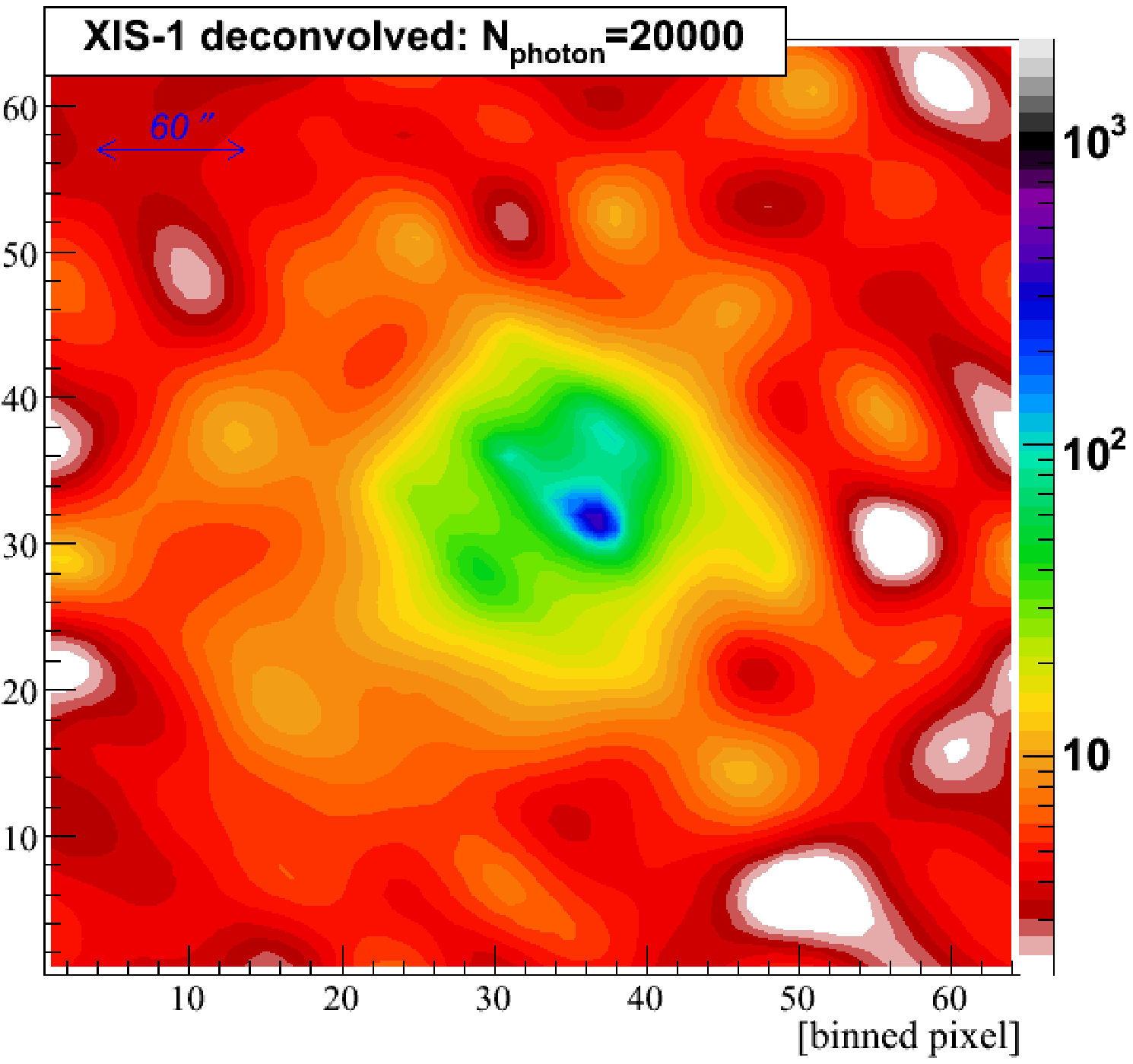}
\FigureFile(4.0cm,){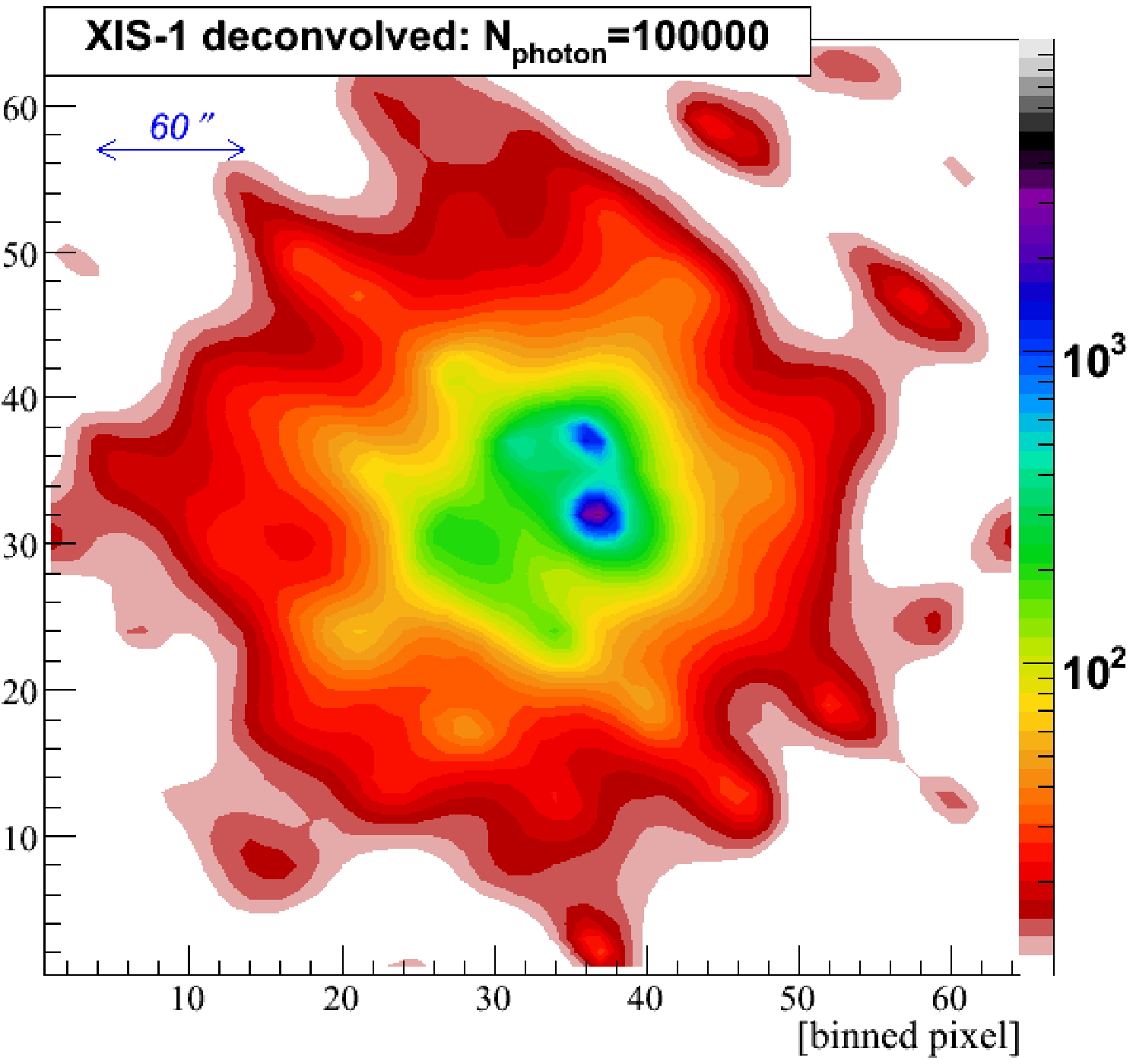}
\FigureFile(4.0cm,){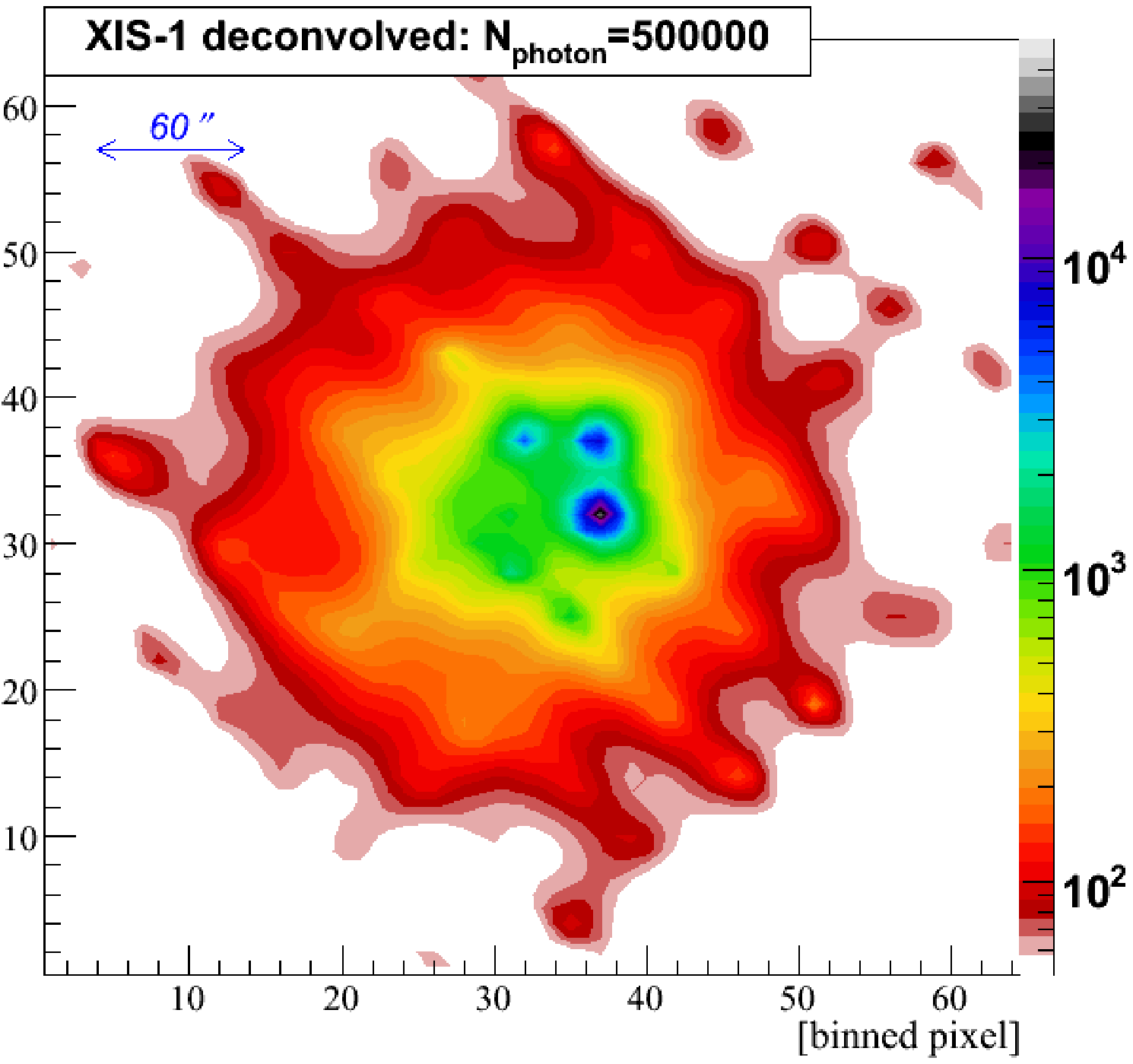}
\end{minipage}
\begin{minipage}[t]{4cm}
\FigureFile(4.0cm,){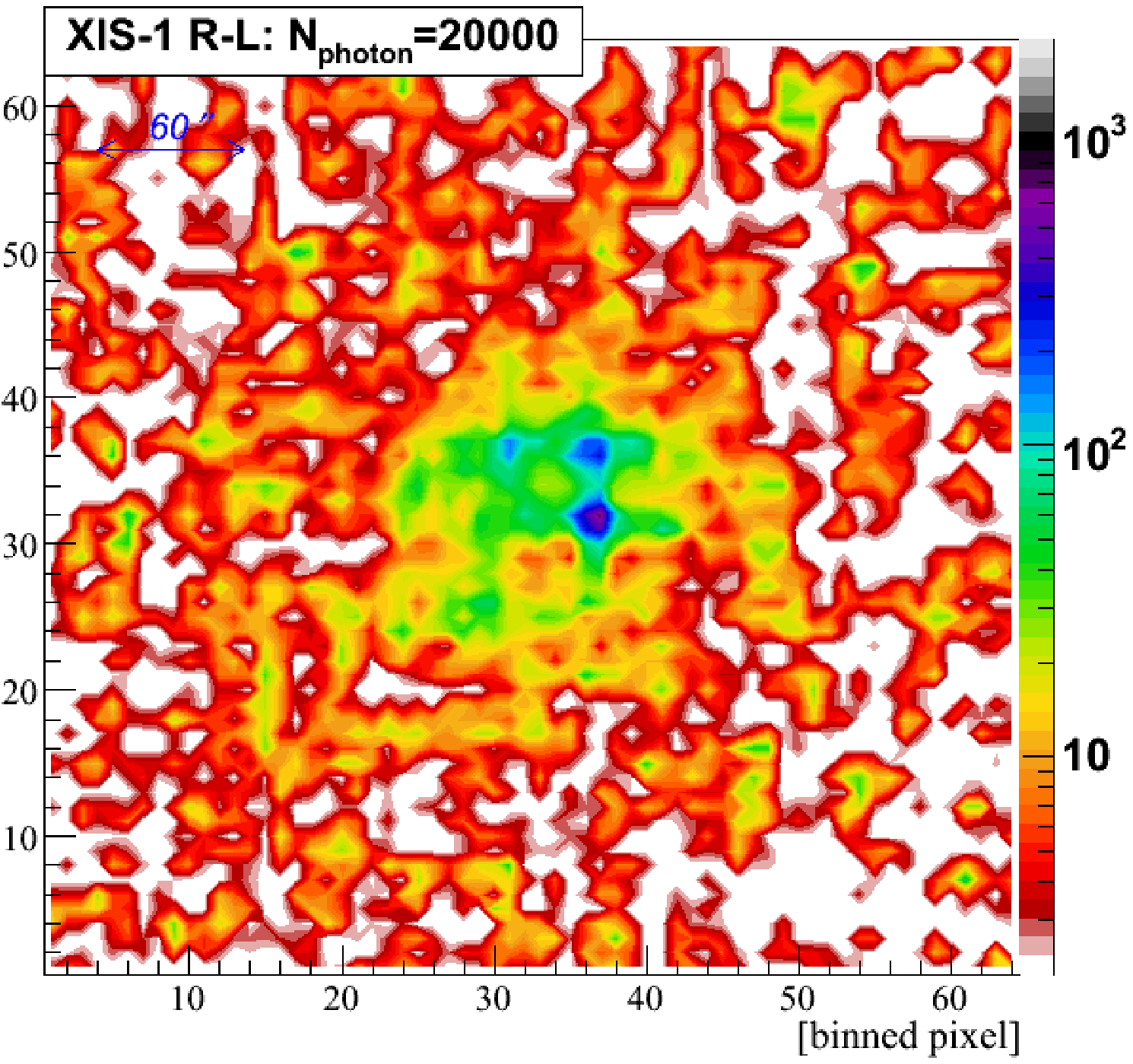}
\FigureFile(4.0cm,){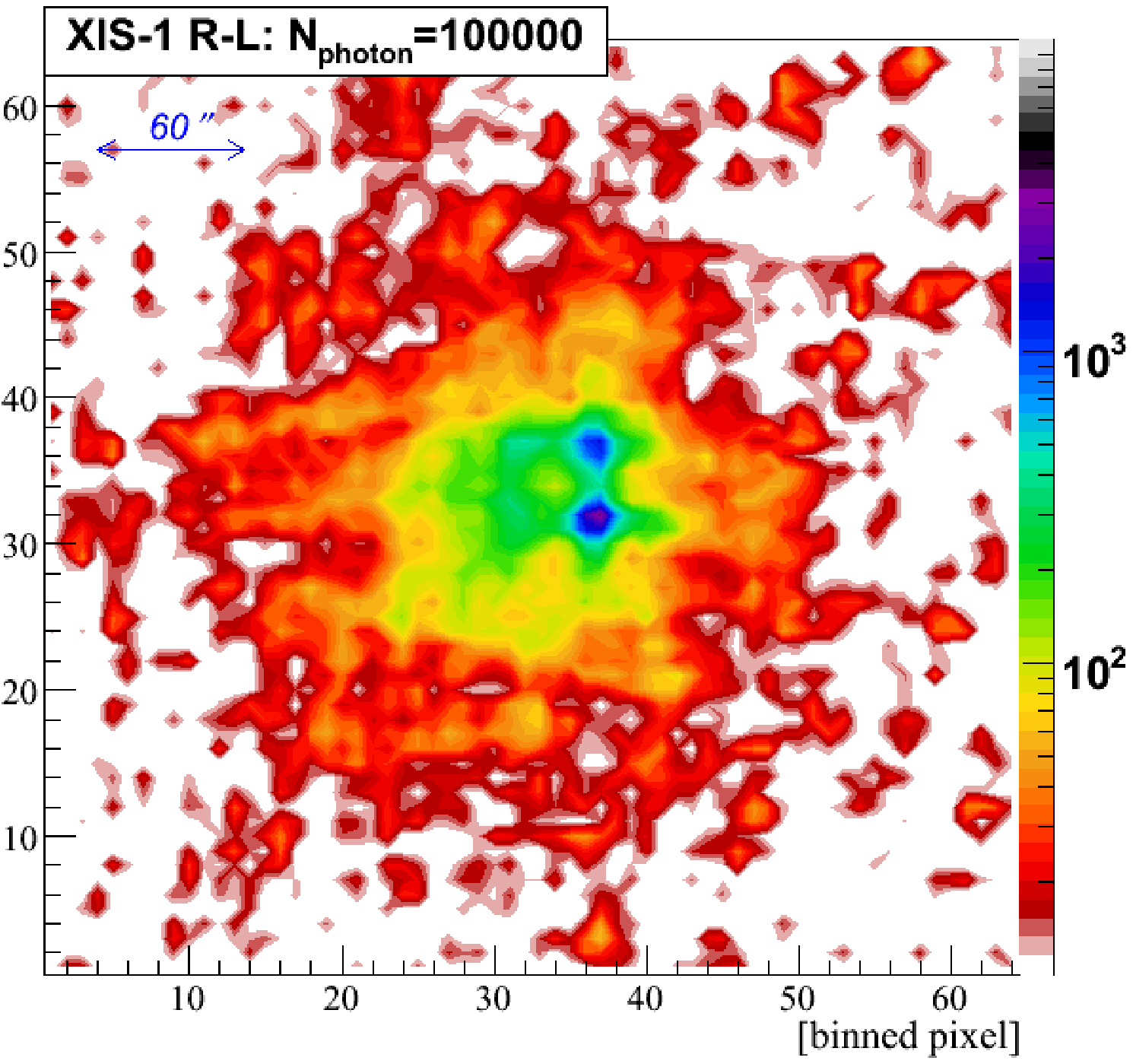}
\FigureFile(4.0cm,){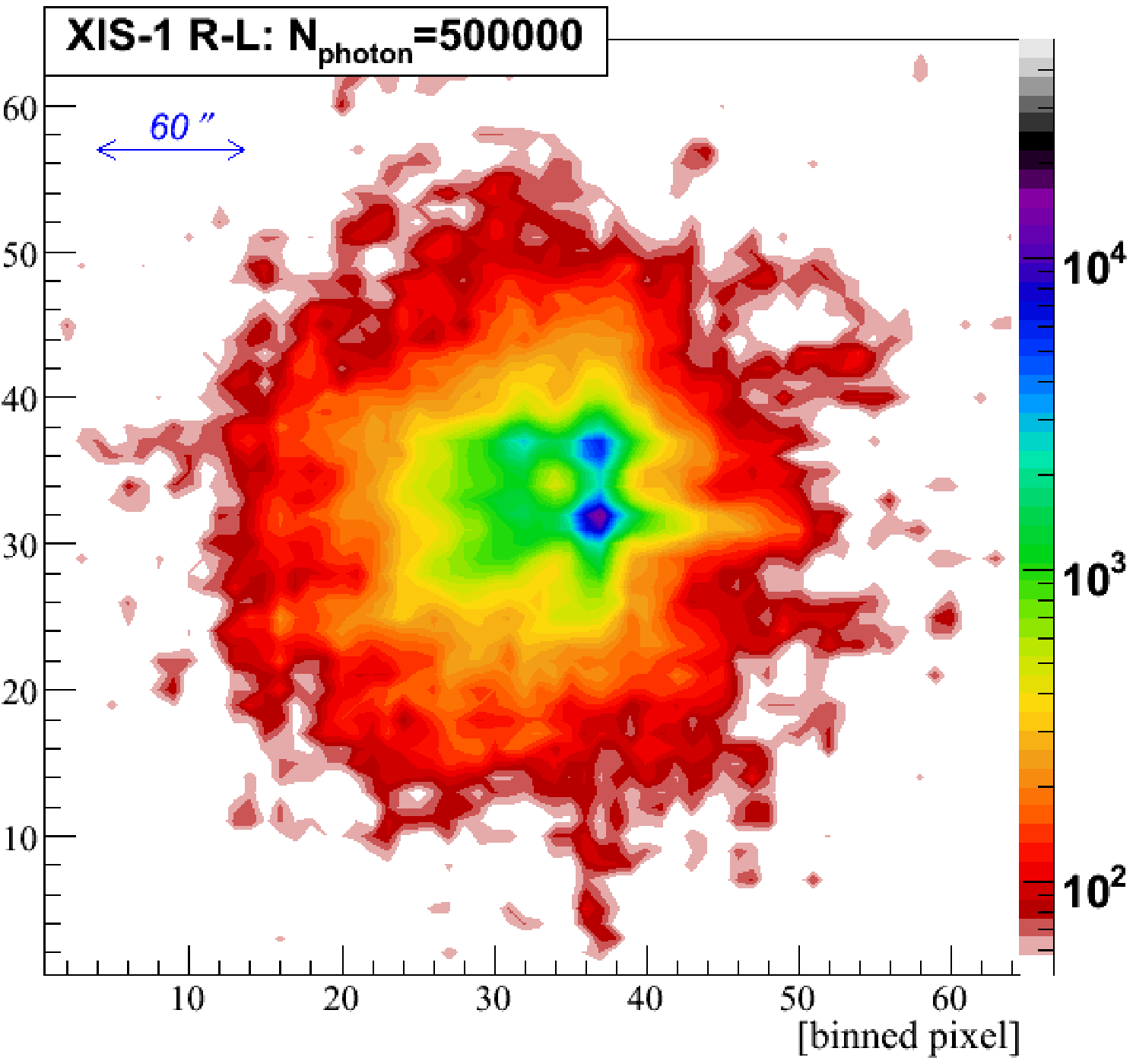}
\end{minipage}

\begin{center}
\caption{

Deconvolution of simulated XIS-1 images.

(Left) Model image consists of three point sources and an extended 
emission.
Relative fluxes of the three point sources (A, B, C) and extended emission
are $4:2:1:40$.

(Center left) Simulated XIS-1 images with photon counts of 20,000,
100,000, 500,000 from top to bottom.  Background photons with Poisson
distribution expected for a nominal blank-sky in a 50-ks exposure are
included.

(Center right) 
Deconvolved XIS-1 images with 20,000, 100,000, and 500,000 photons
from top to bottom by the present method.

(Right) Deconvolved XIS-1 images by a Richardson-Lucy method
after 100 iterations.

}
\label{fig:simdecon}

\end{center}
\end{figure}

We have tested the fidelity of the present deconvolution method 
using simulated XIS-1 images.  The simulated image consists 
of three point sources and an extended source as shown in 
figure \ref{fig:simdecon}. The surface brightness of the extended emission 
is represented by a simple $\beta$ model:
\begin{equation}
S(r)=S_0\left\{1+(r/r_c)^2\right\}^{-3\beta+0.5},
\end{equation}
where the critical radius is set to $r_c=30''$ and $\beta$ to 0.5.
Relative fluxes of three point sources and extended emission
are set at $4:2:1:40$.
To study how fidelity depends on photon statistics, we have 
deconvolved three images containing total signal photons of 20,000, 100,000, 
and 500,000 and background photons expected for 
nominal blank-sky in a 50-ks exposure
\citep{Mitsuda2007}. 
These images are then compared with images deconvolved by a Richardson-Lucy 
method with 100 iterations in figure \ref{fig:simdecon}.

One can see that the Richardson-Lucy method adds photons to high points in 
Poisson fluctuation noise and depletes photons from low 
points in the fluctuation. Our method, on the other hand, smooths out 
the fluctuation below the predetermined signal-to-noise ratio 
while reproducing the three point sources and the extended source 
well. Such noise filter can be added to the Richardson-Lucy method 
to suppress artifacts but the filtering strategy is strongly 
coupled with the number of iteration. 
We note that the raw image of 100,000 photons
at the middle row in Figure  \ref{fig:simdecon}
has about $\sim 50$ signal photons 
and $\sim 3$ noise photons per tile in average within
the critical radius $r_c=30''$.

\subsection{Cen A}

The data of Cen A are divided into two energy bands, $>3$~keV and $<3$~keV, 
and the two images are deconvolved separately.  In the hard band, 
the flux of Cen A is known to be dominated by the central point source
\citep{Evans2004,Markowitz2007}.  
In the soft band, an extended jet profile with a scale of
$30''$--$180''$ has been observed by Chandra
\citep{Kraft2002,Kataoka2006}.

\subsubsection{Hard band: 3--10 keV}
\label{sec:cena_hard} 

\begin{figure}
\begin{center}

\FigureFile(5.5cm,){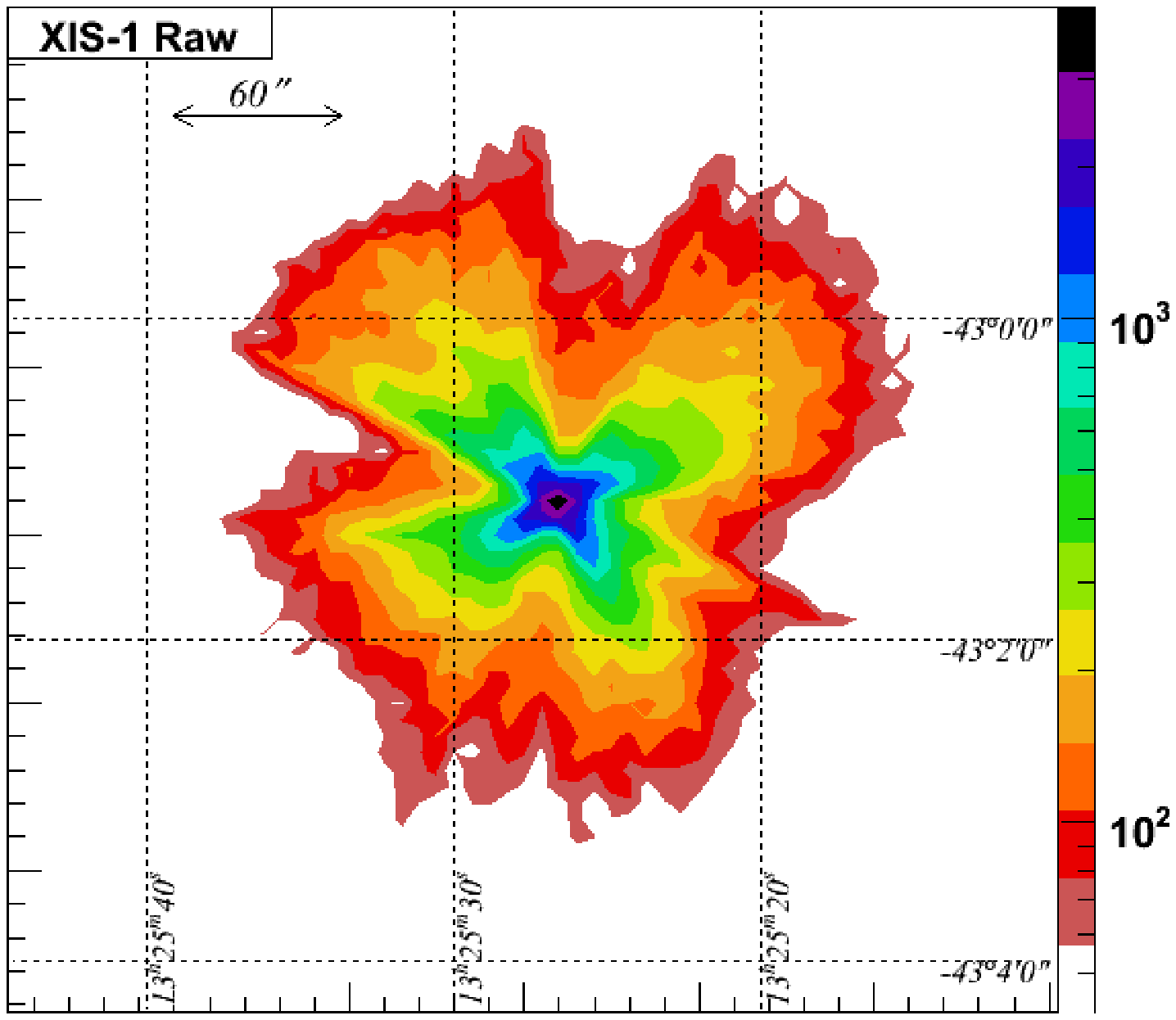}
\FigureFile(5.5cm,){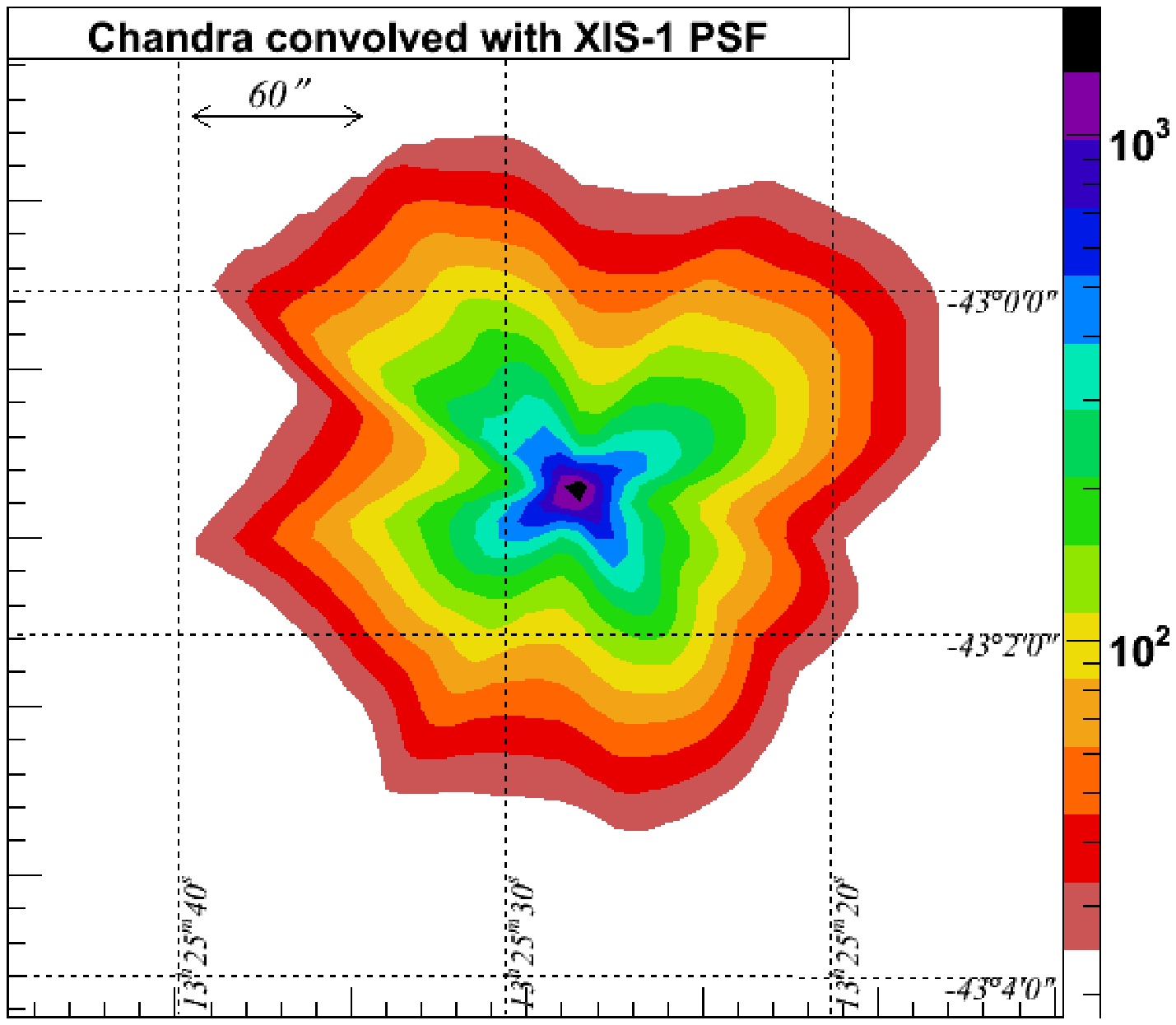}

\FigureFile(5.5cm,){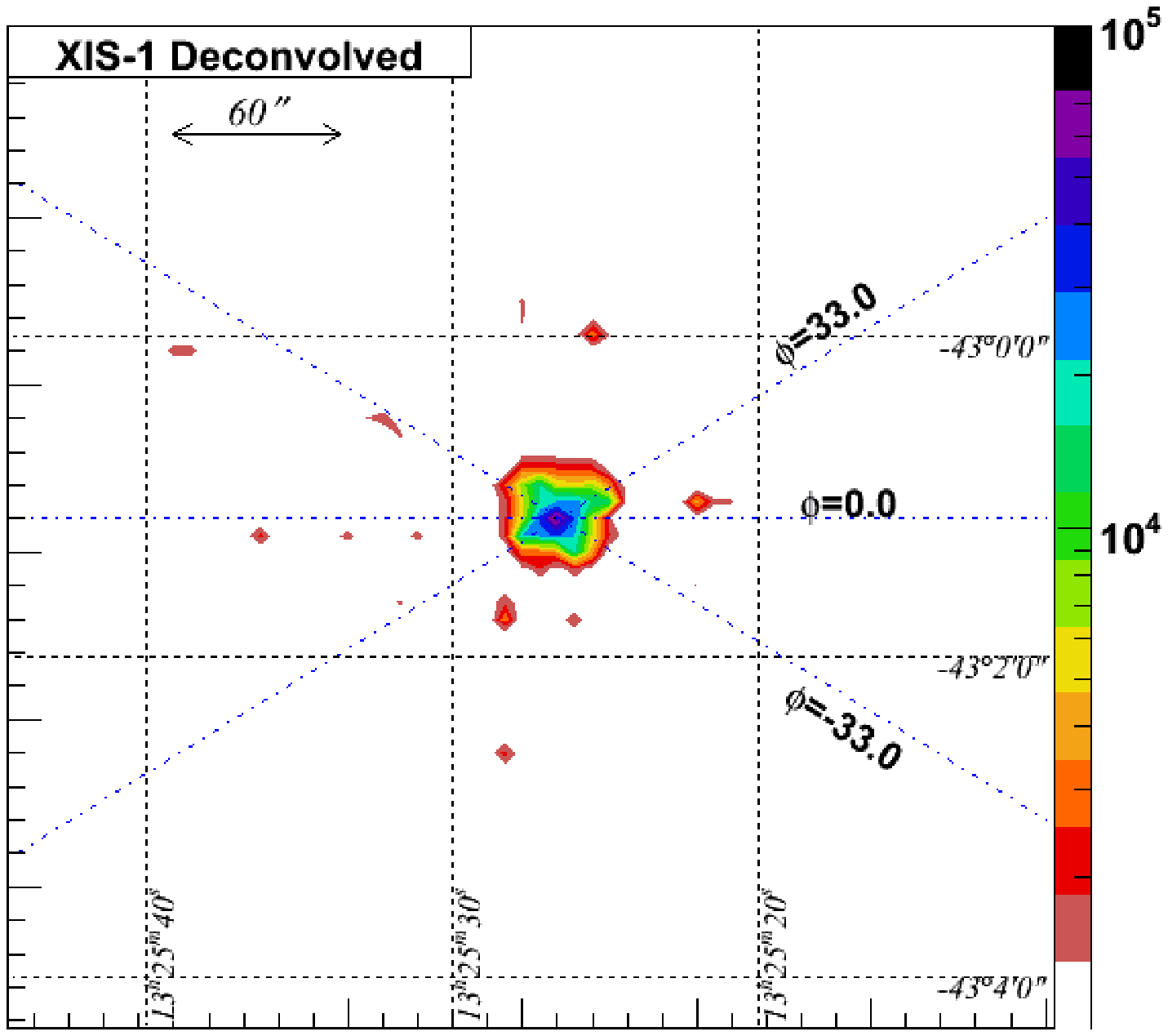}
\FigureFile(5.5cm,){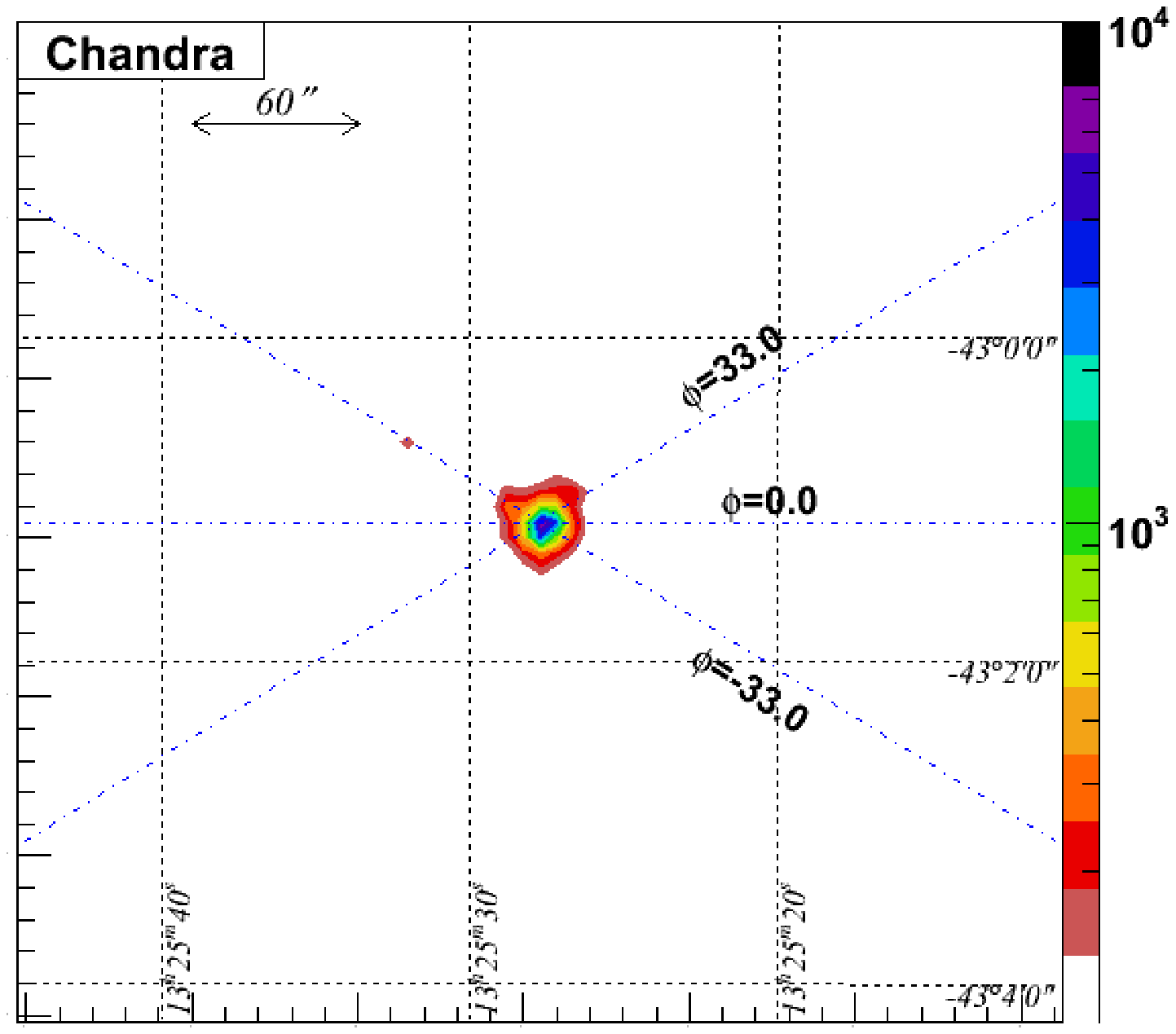}

\caption{ 
  XIS-1 raw image ({\it top left}), 
  convolved Chandra ACIS image with XIS-1 PSF ({\it top right}), 
  XIS-1 deconvolved image ({\it bottom left}), 
  and Chandra ACIS image ({\it bottom right}) of Cen A in 3--10 keV.
  All images are binned with a same unit tile size of $\sim
  6''\times 6''$.  
  Contour colors are spaced logarithmically.  
  Notice that
  the Chandra ACIS image suffers from the pile-up effect at the
  nucleus of Cen A (the brightest point).  
  Blue dash-dot lines 
  indicate the sliced directions in figure \ref{fig:cena_hard_slx}.  
}
\label{fig:cena_hard_img}
\end{center}
\end{figure}

Figure \ref{fig:cena_hard_img} shows the observed image of XIS-1 in 
3-10 keV band, its deconvolved image, and the Chandra ACIS image
rebinned to the tile size. We have also convolved 
the Chandra image with the XIS-1 PSF and shown in the top right panel.
We note that the Chandra ACIS image is distorted by the pile-up effect 
at the Cen-A nucleus and the one shown in figure \ref{fig:cena_hard_img} 
has been corrected for the distortion by using the Suzaku XIS image.

\begin{figure}
\begin{center}

\FigureFile(5.5cm,){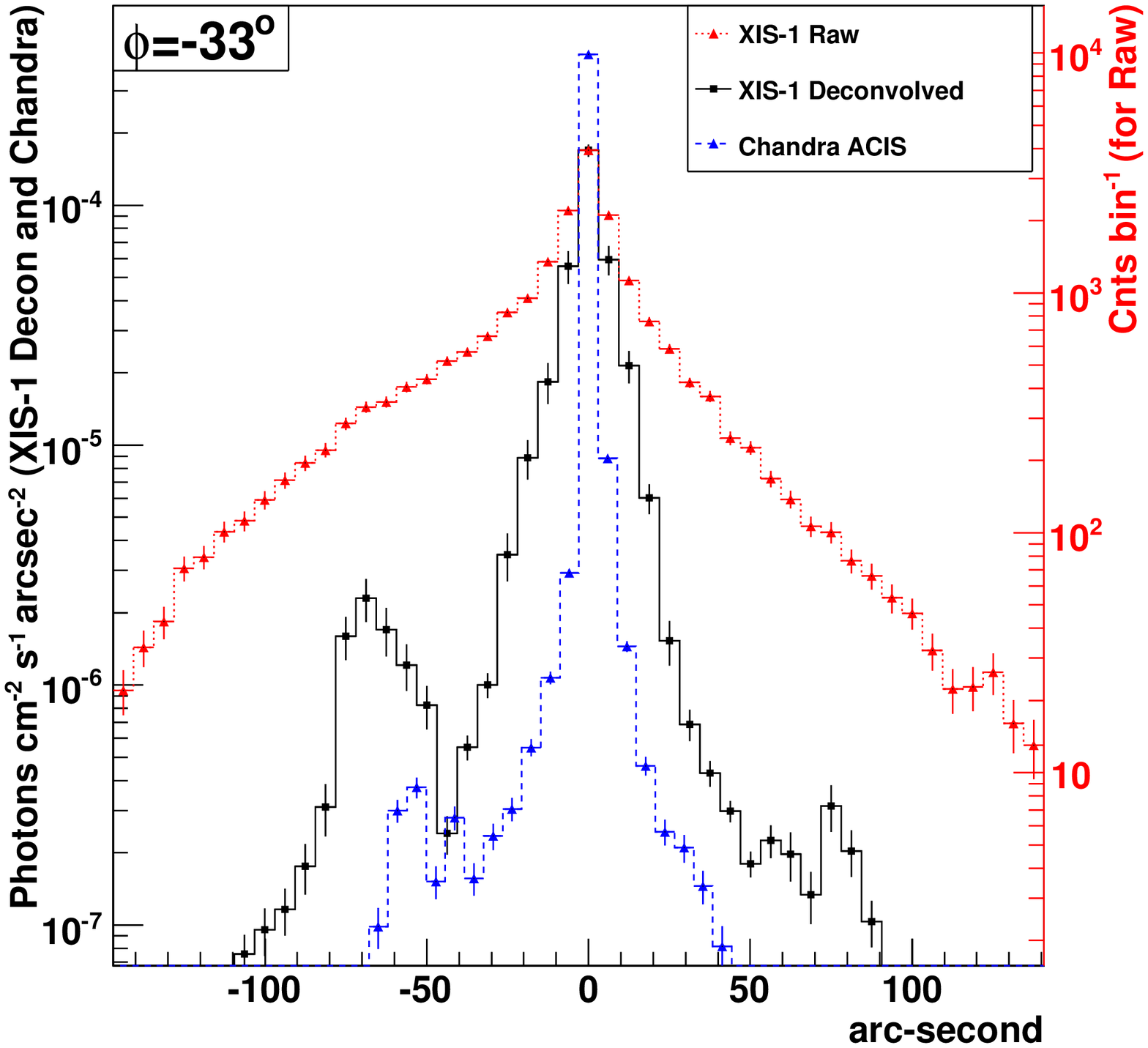}
\FigureFile(5.5cm,){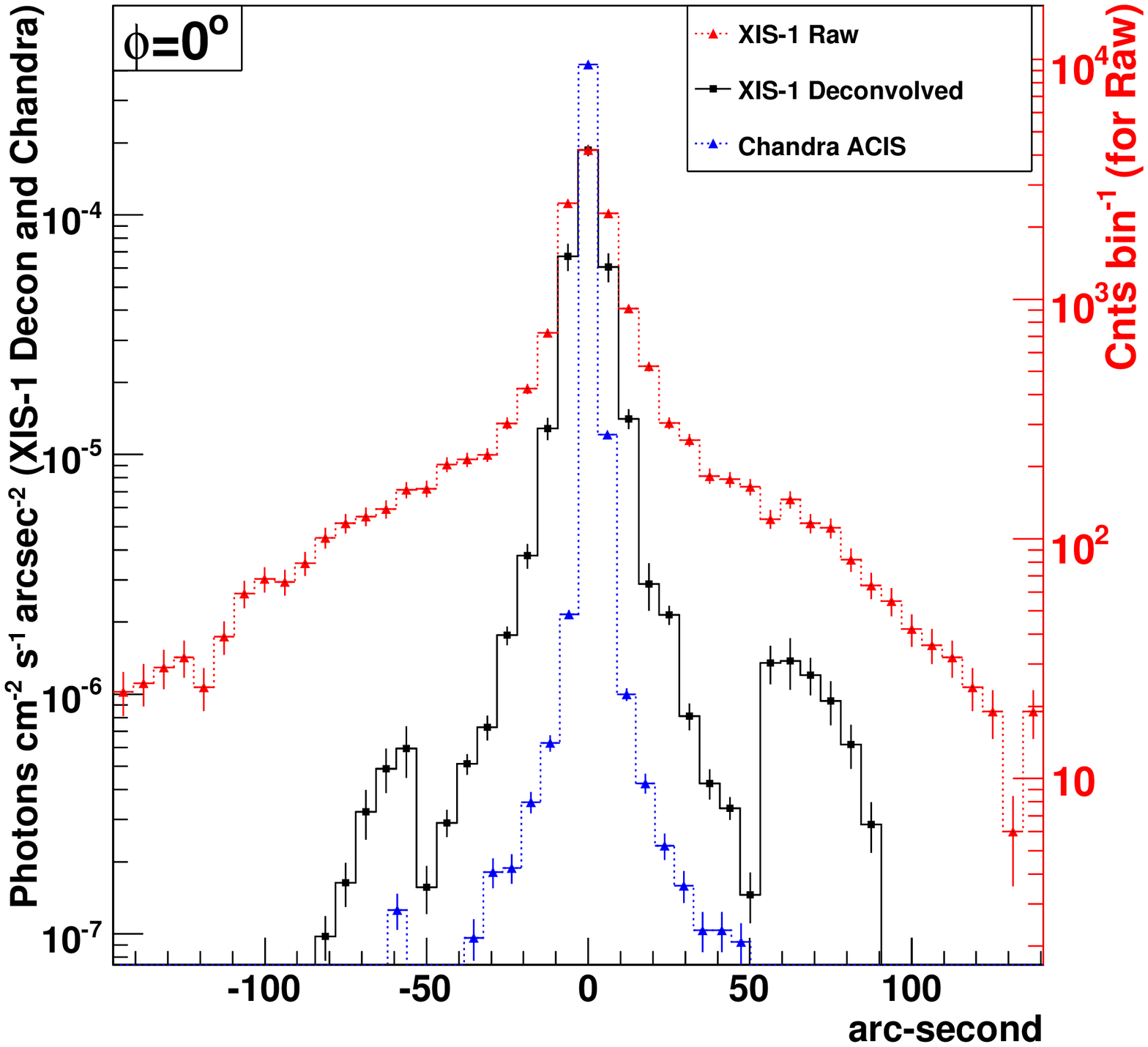}
\FigureFile(5.5cm,){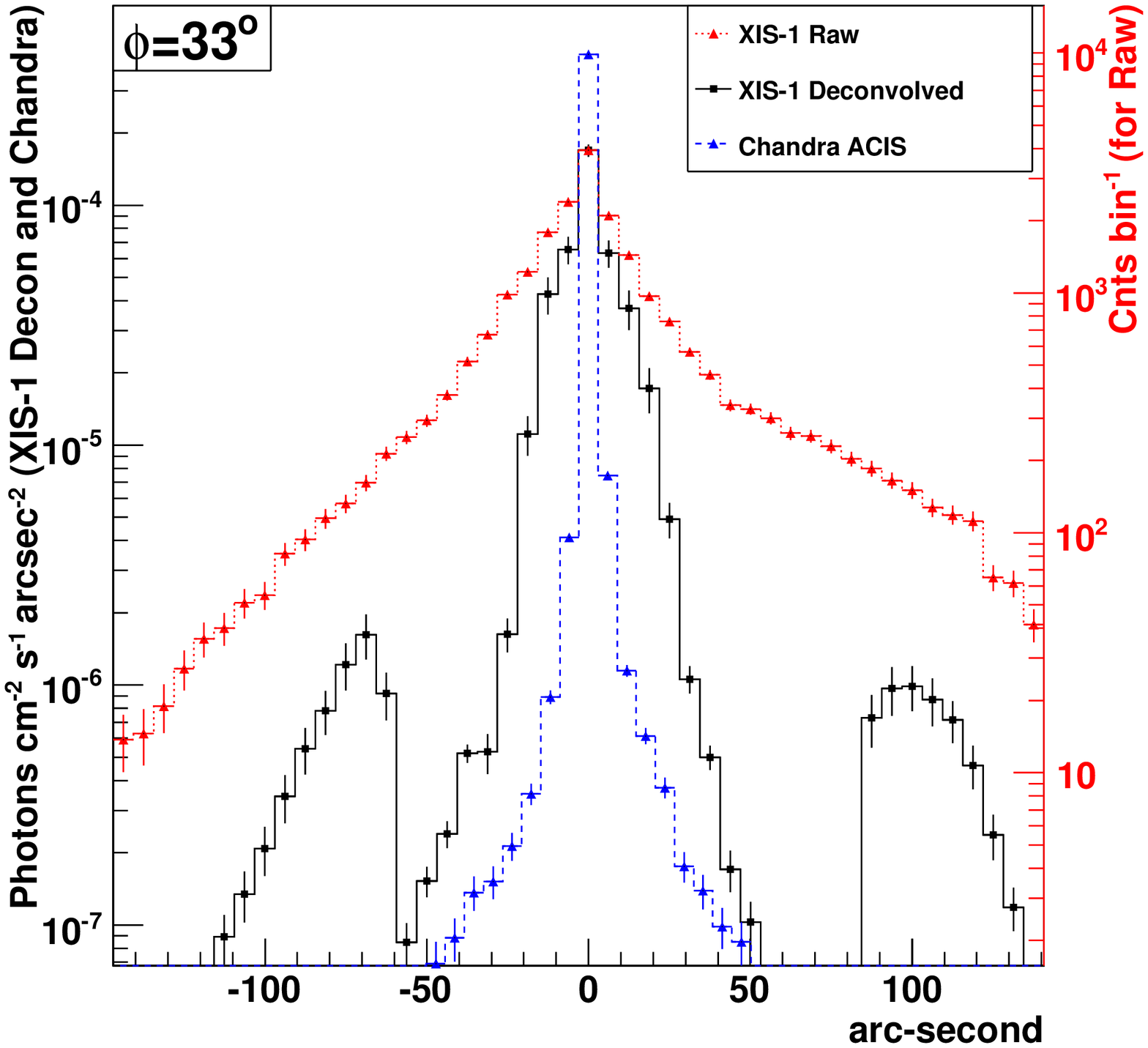}
\caption{ 
  Cross-section profiles of the XIS-1 raw image (red dot) 
  deconvolved image (black solid) and Chandra ACIS image (blue dash) of Cen A
  in 3-10 keV band,
  sliced along the blue dash-dot lines in figure \ref{fig:cena_hard_img}.
  Error bars represent 1-$\sigma$ photon-statistics uncertainties.
  The vertical scale of the raw image is normalized by the peak value
  to the deconvolved image.
  Notice that the pile-up effect at the peak of the Chandra ACIS image 
  (one pixel at the center)
  is corrected.
}
\label{fig:cena_hard_slx}

\end{center}
\end{figure}

We have studied the deconvolved Suzaku image by slicing it along 
the 3 blue dash-dot lines drawn at $\phi = -33^\circ, 0^\circ$, and 
$33^\circ$ from the Right Ascension axis crossing the Cen A core 
in figure \ref{fig:cena_hard_img}.  
The surface brightness profiles are compared with 
those of the XIS-1 raw image and the Chandra ACIS image in figure
\ref{fig:cena_hard_slx}.  
The surface brightness was calculated by assuming a power-law spectrum 
with photon index $\Gamma =1.8$ \citep{Markowitz2007}.  The vertical
scale of the XIS-1 raw image was normalized at the peak to the
deconvolved image.  

In the deconvolved image, the large wings extending over $\sim 120''$ 
in the raw images are drastically reduced.  The width
of the central peak in the deconvolved image is wider by $12''-24''$
than the ACIS image.  This is interpreted that the spatial resolution 
has been improved to $12''-24''$ by deconvolution but not better.  
The spatial resolution depends on the azimuth angle: It is the worst 
at $\phi=33^\circ$, which corresponds to the
direction along which the PSF is extended.  We find two artifact peaks 
in the region $50''-100''$ away from the peak 
in all three profiles at about 1/50 
of the peak surface brightness.  This shows limitation of the present 
deconvolution method. With the present modeling of the XIS PSF, 
our deconvolution reconstructs the image down to about 1:50. 
A third exponential component with a wider wing may be 
required to improve the dynamic range.  

\subsubsection{Soft band: 0.5--3 keV}
\begin{figure}
\begin{center}

\FigureFile(5.5cm,){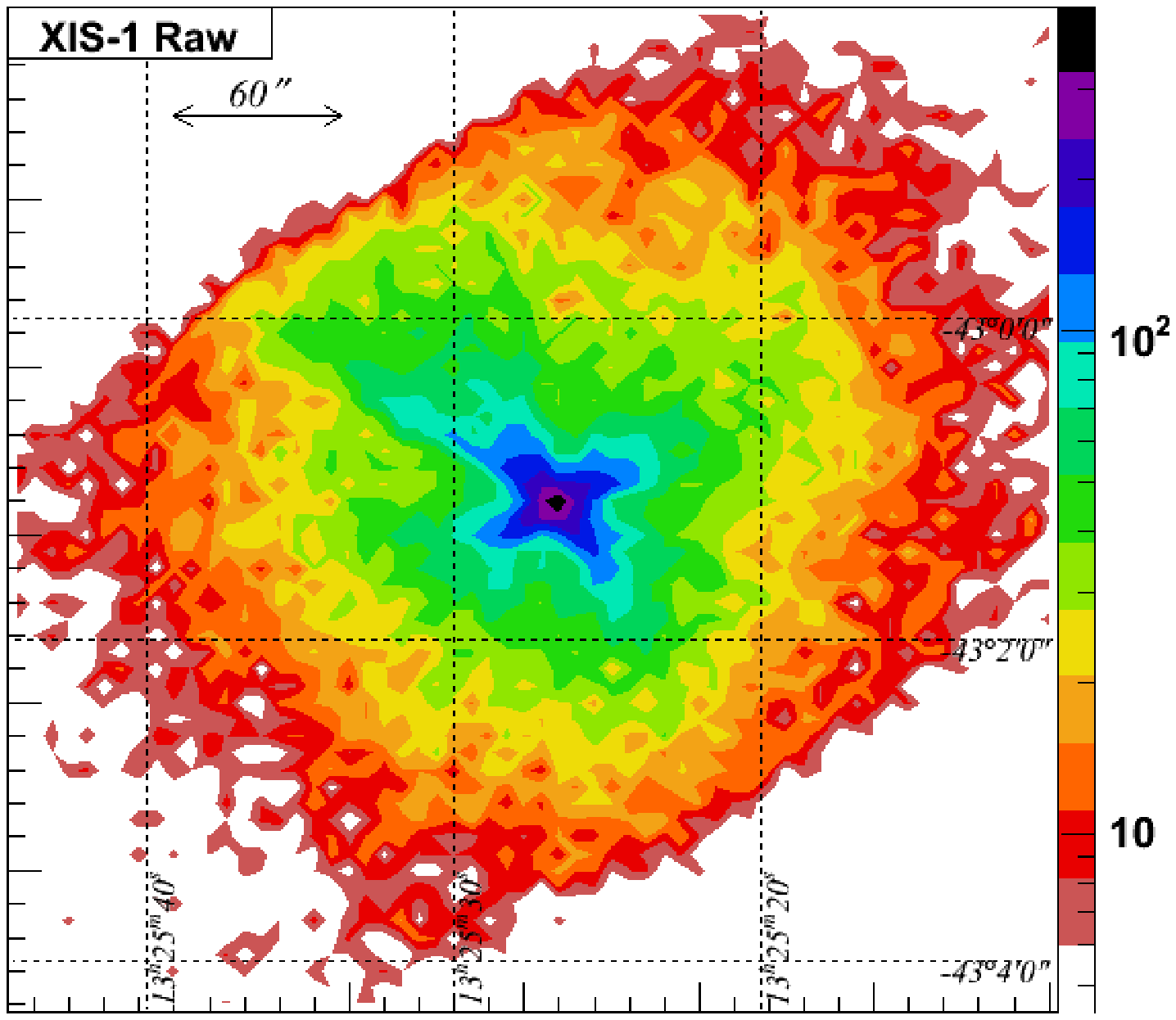}
\FigureFile(5.5cm,){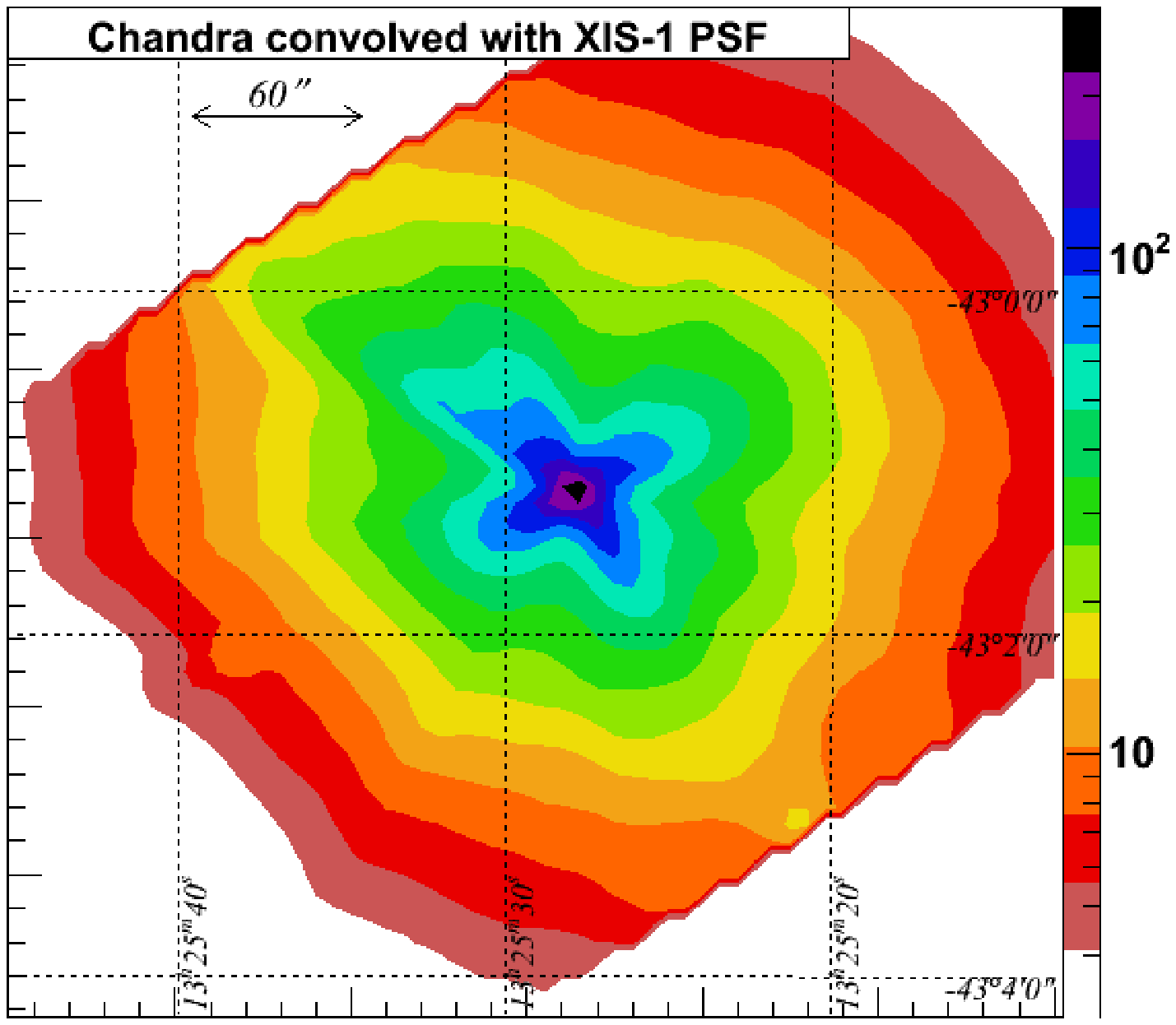}

\FigureFile(5.5cm,){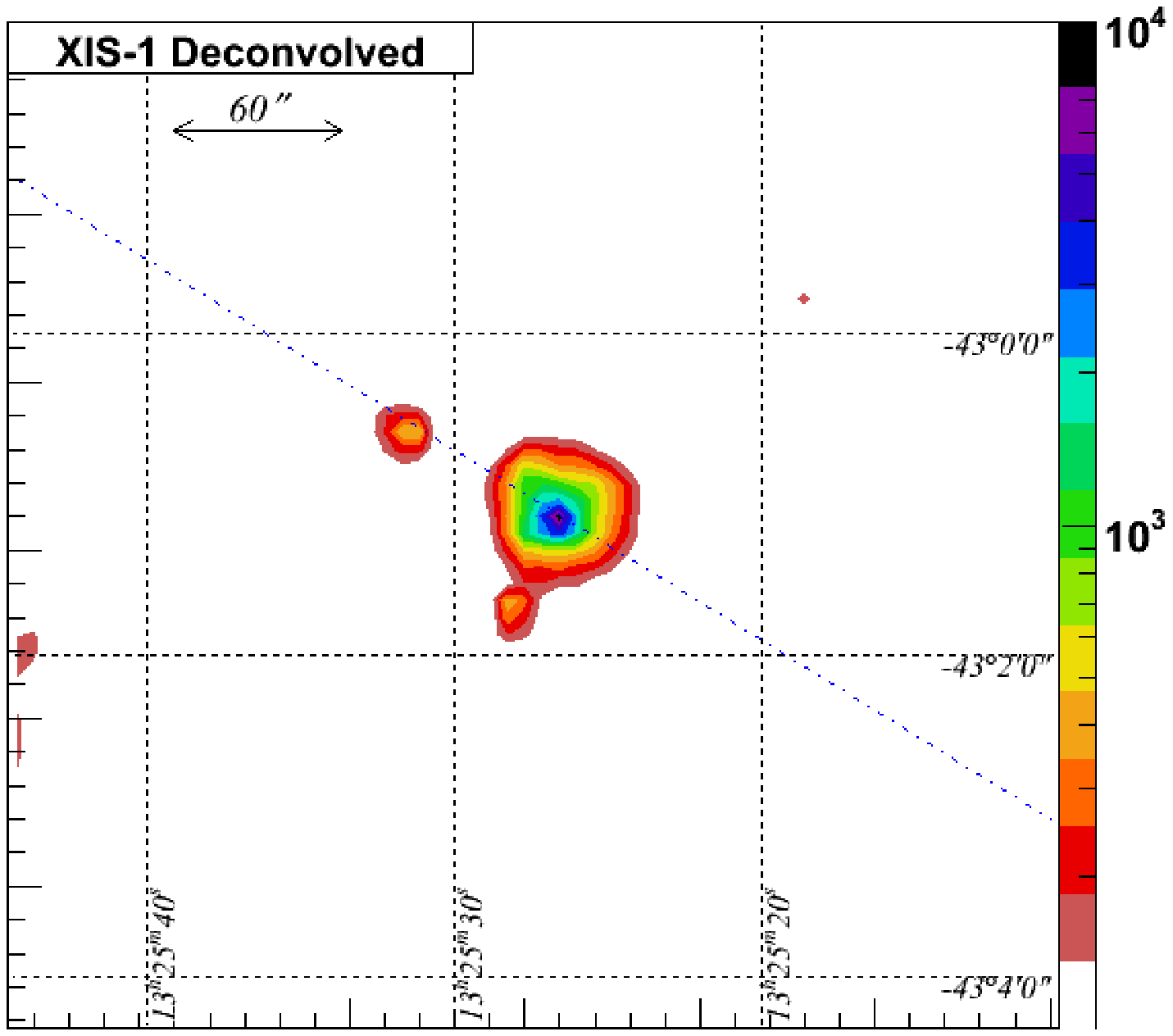}
\FigureFile(5.5cm,){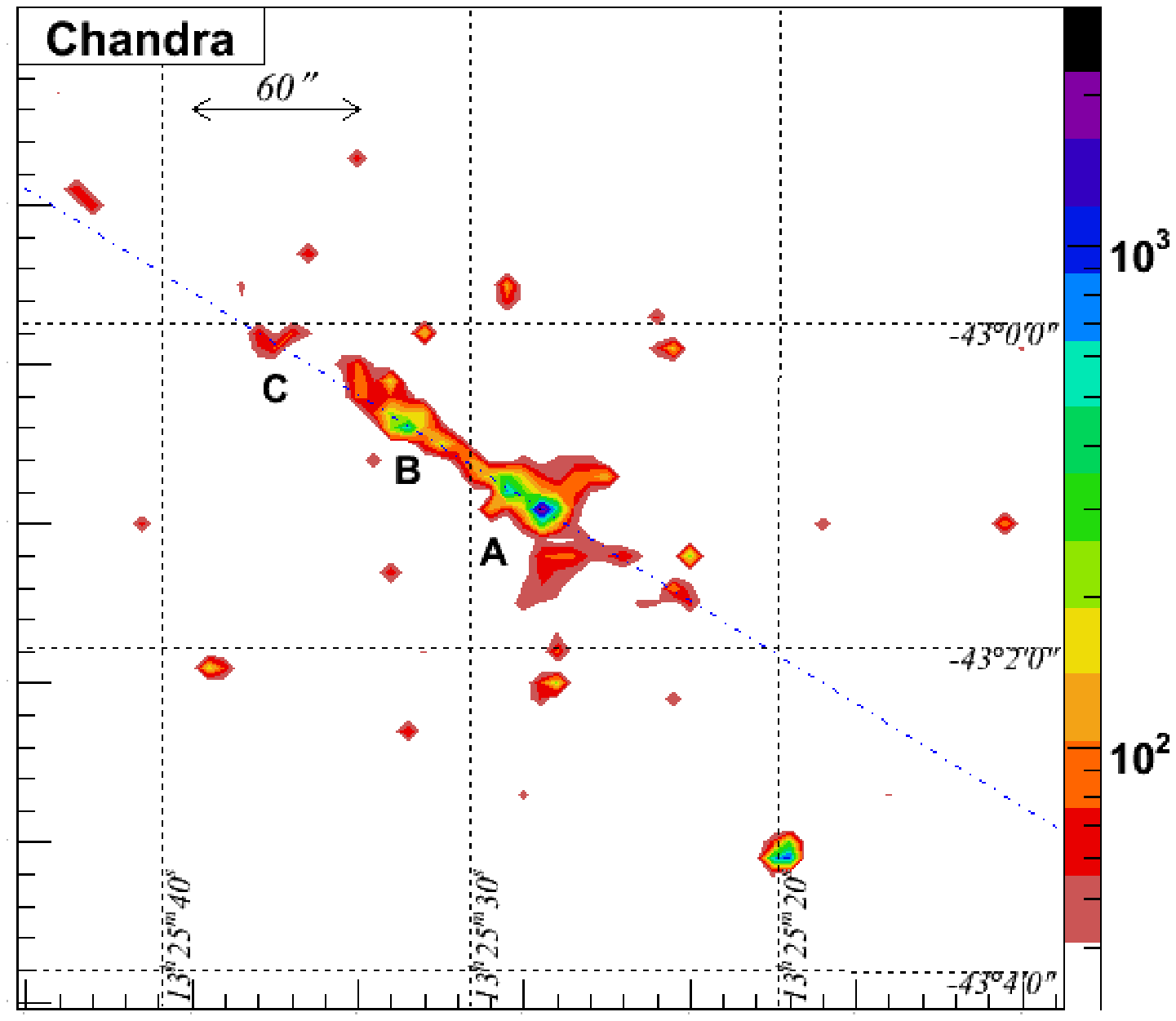}

\caption{ 
  XIS-1 raw image ({\it top left}), 
  convolved Chandra ACIS image with XIS-1 PSF ({\it top right}), 
  XIS-1 deconvolved image ({\it bottom left}), 
  and Chandra ACIS image ({\it bottom right}) of Cen A in 0.5--3 keV band.
  All images are binned with a same unit tile size of $\sim
  6''\times 6''$.  
  In images of XIS-1, the blank
  areas at the top left and the bottom right are not covered
  by clocked CCD window in this observation.  
  Note that the peak of the Chandra ACIS image (one pixel at the center)
  suffering from a pile-up effect 
  has been corrected using the flux measured by the Suzaku XIS.
  The blue dash-dot lines on the XIS-1 deconvolved image and 
  the Chandra image
  indicate the sliced direction in Figure \ref{fig:cena_soft_slx}.
  The labels ``A'', ``B'', ``C'' in the Chandra image identify the peaks.
}

\label{fig:cena_soft_img}
\end{center}
\end{figure}

Figure \ref{fig:cena_soft_img} shows the observed image of XIS-1 
in 0.5-3 keV band, the deconvolved image, and the 
rebinned Chandra ACIS image.
The convolved ACIS image with the XIS-1 PSF
is also shown in the top right panel.
The pile-up effect at the nucleus of Cen A
in the ACIS image is corrected 
from the flux measured by the XIS.
Our deconvolution reconstructs the bright blob ``B'' but not the 
blob ``C'' in the jet.

\begin{figure}
\begin{center}
\FigureFile(7.5cm,){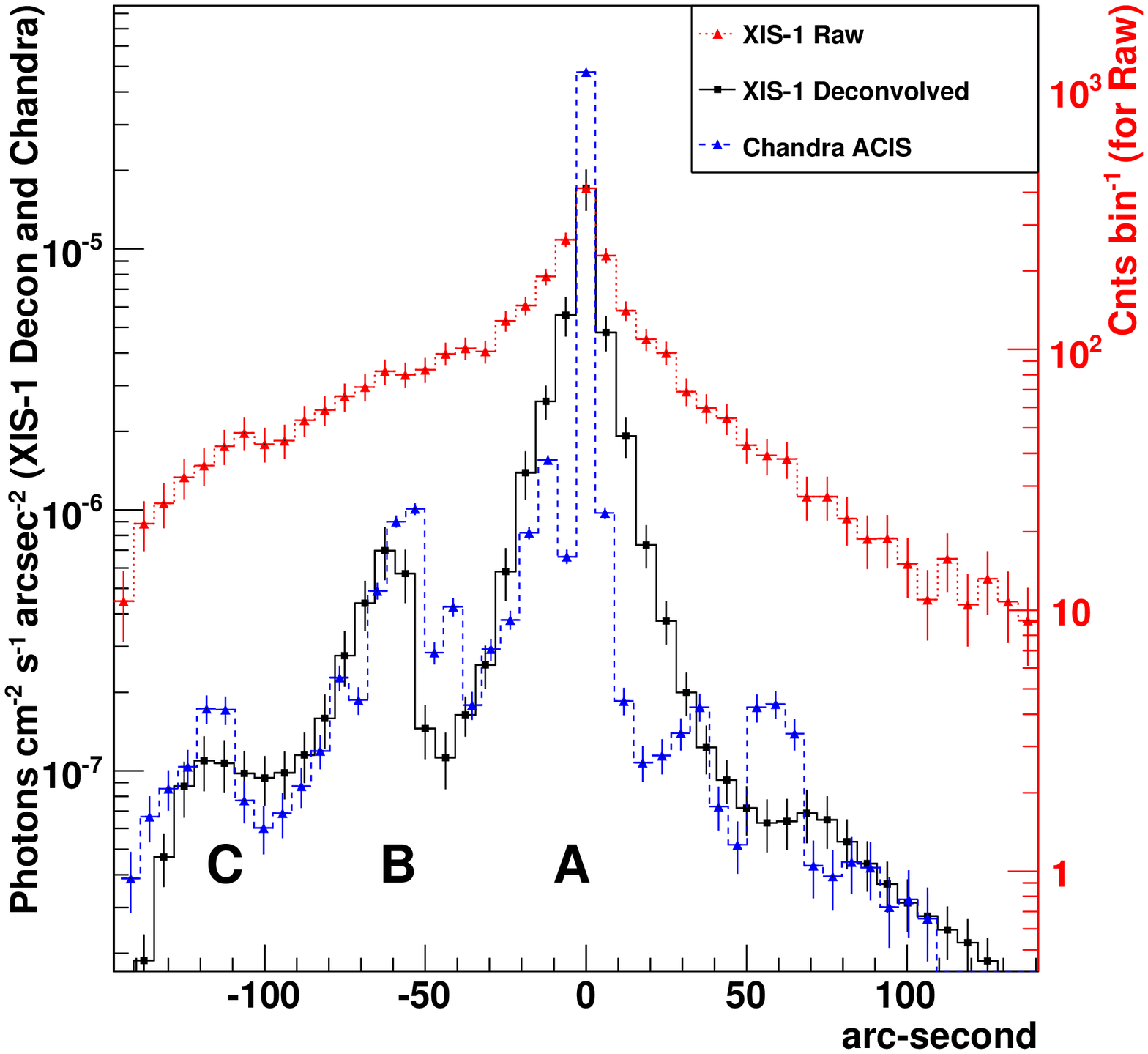}
\caption{ 
  Cross-section profiles of XIS-1 raw image (red dot) 
  and deconvolved image (black solid)
  and Chandra ACIS image (blue dash)
  of Cen A in 0.5-3 keV sliced along the direction of the jet
  (blue dash-dot lines in figure \ref{fig:cena_soft_img}).
  Error bars represent 1-$\sigma$ photon-statistics uncertainties.
  The vertical scale of XIS-1 raw data is normalized at the peak value
  to the deconvolved image.
  The peak of the Cen-A nucleus in the Chandra profile
  has been corrected for the pile-up effect.
  The labels ``A'', ``B'', ``C'' indicate the peaks in 
  figure \ref{fig:cena_soft_img}.
}

\label{fig:cena_soft_slx}
\end{center}
\end{figure}

Figure \ref{fig:cena_soft_slx} shows the cross-section profiles of the
XIS-1 raw, deconvolved, and ACIS images sliced along the direction to
the north-east jet (a dash-dot blue line in figure
\ref{fig:cena_soft_img}).  Here, the surface brightness is calculated
assuming a power-law spectrum with photon index
$\Gamma =1.3$ \citep{Markowitz2007}.  
The profile around peak is well restored in the deconvolved image
with a spatial resolution of $20''$.  
The restored image is consistent with that by Chandra
if we take the pile-up effect in the Chandra image
into account.

\subsection{PSR B1509-58 and RCW 89}

\begin{figure}
\begin{center}

\FigureFile(8cm,){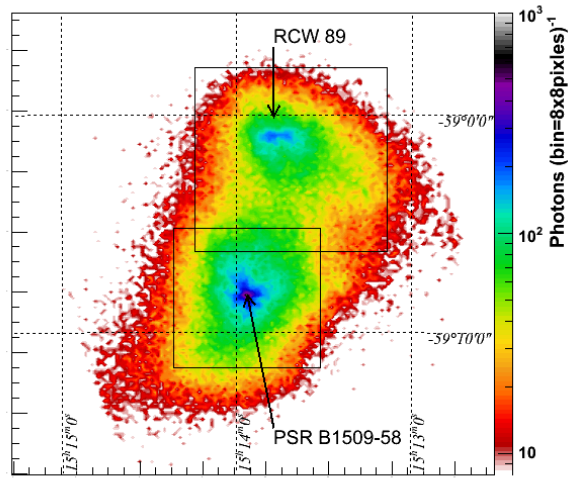}
\caption{ 
XIS-1 overall raw image of PSR B1509-58 and RCW 89  
in 0.5--3 keV band.
Solid boxes represent the regions of which images 
are applied to the deconvolution analysis.
}
\label{fig:psr1509raw}
\end{center}
\end{figure}

The observed region of PSR B1509-58 includes a bright point source PSR
B1509-58 and an extended source RCW 89 \citep{Yatsu2005,DeLaney2006}.
Figure \ref{fig:psr1509raw} shows the overall image by XIS-1.  
The hard band image exhibits only one point source, PSR B1509-58, and no 
extended emission. The image in the soft band consists of complex extended 
emissions as shown in the next subsection. 

\subsubsection{PSR B1509-58: 0.5--3 keV}
The raw image of the region including PSR B1509-58 and RCW 89 is shown 
in figure \ref{fig:psr1509raw}. The image includes interesting extended 
emissions and a point source, and can be viewed as a typical target for 
our image deconvolution program. 

\begin{figure}

\FigureFile(5.5cm,){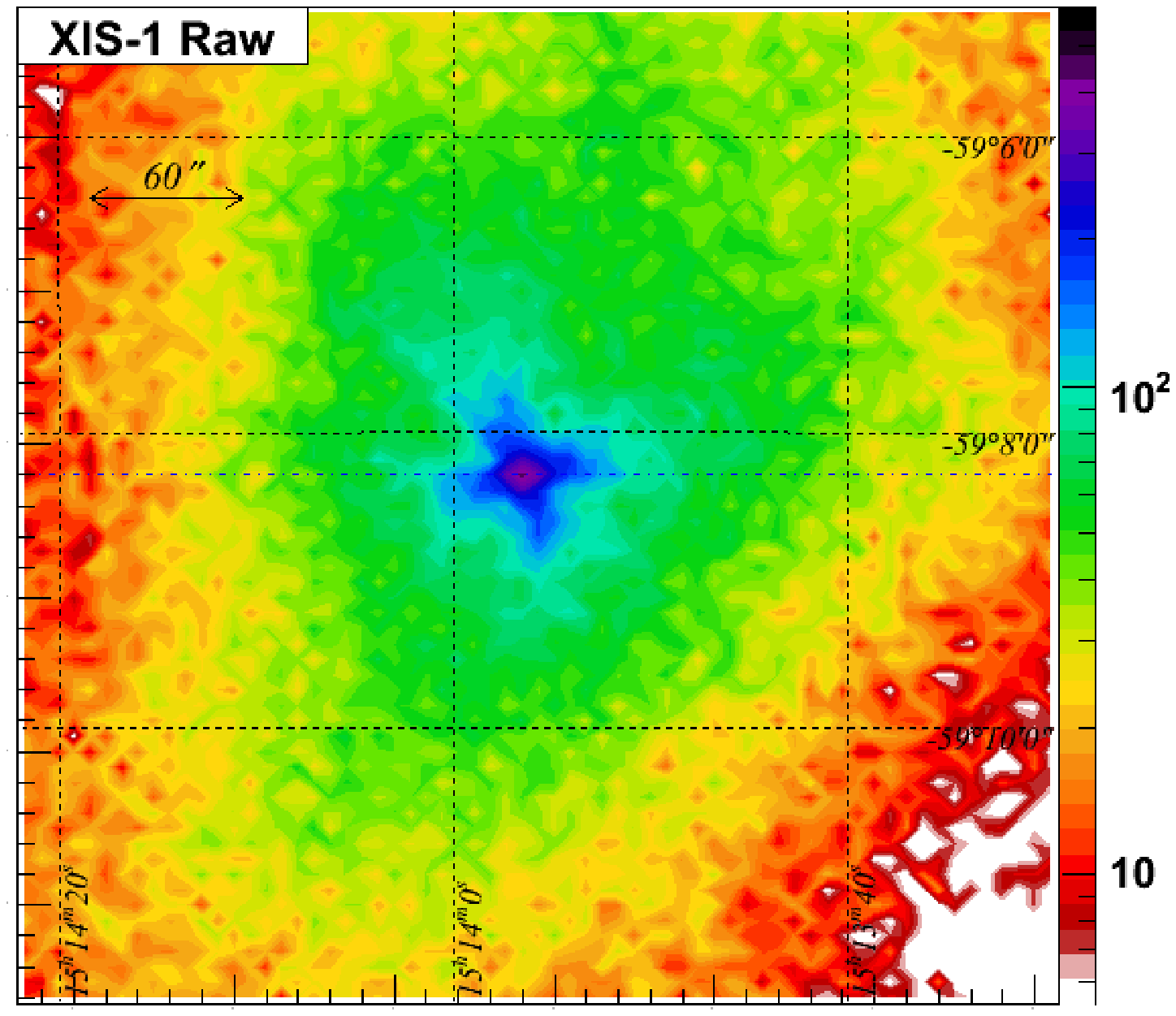}
\FigureFile(5.5cm,){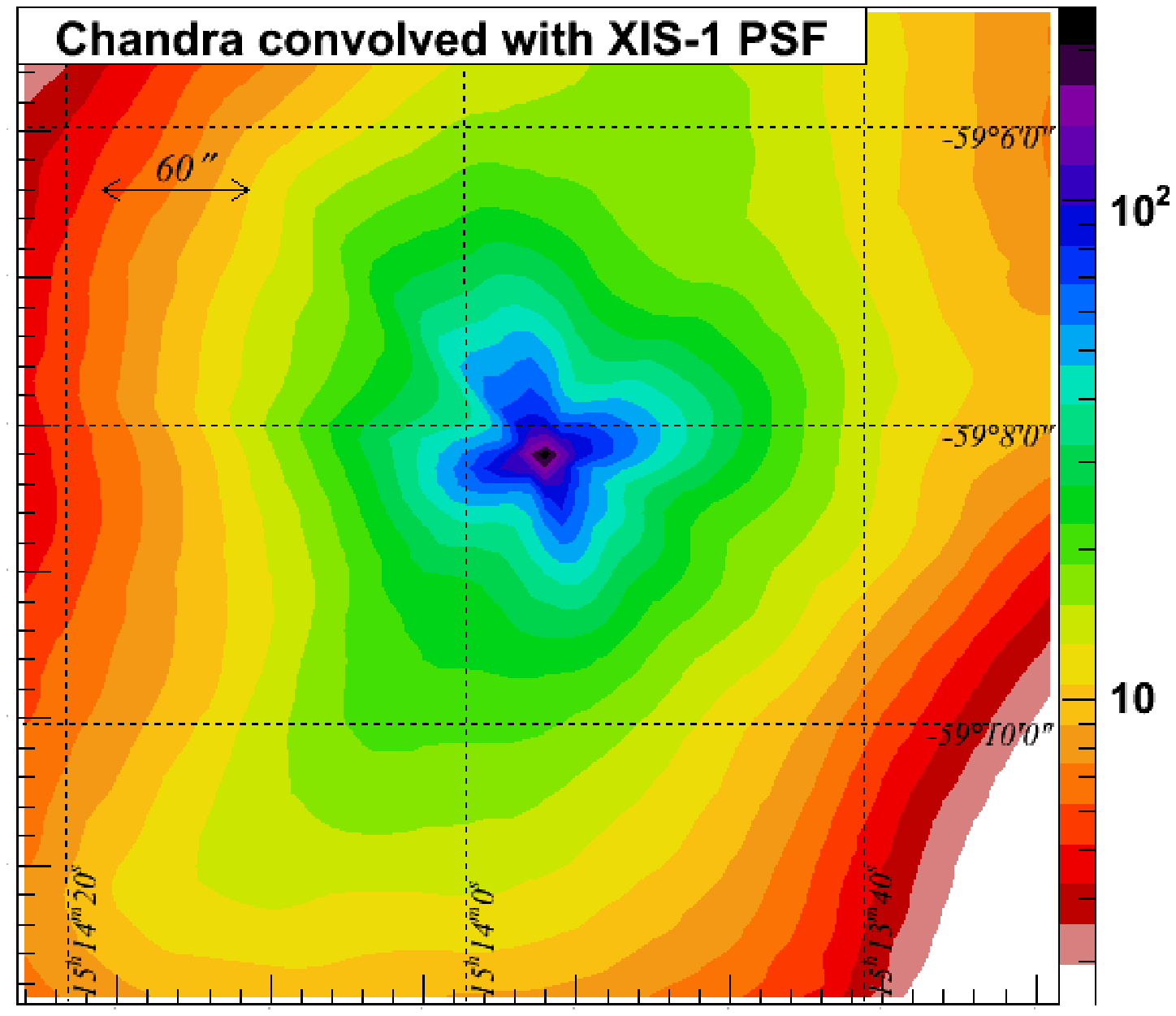}

\FigureFile(5.5cm,){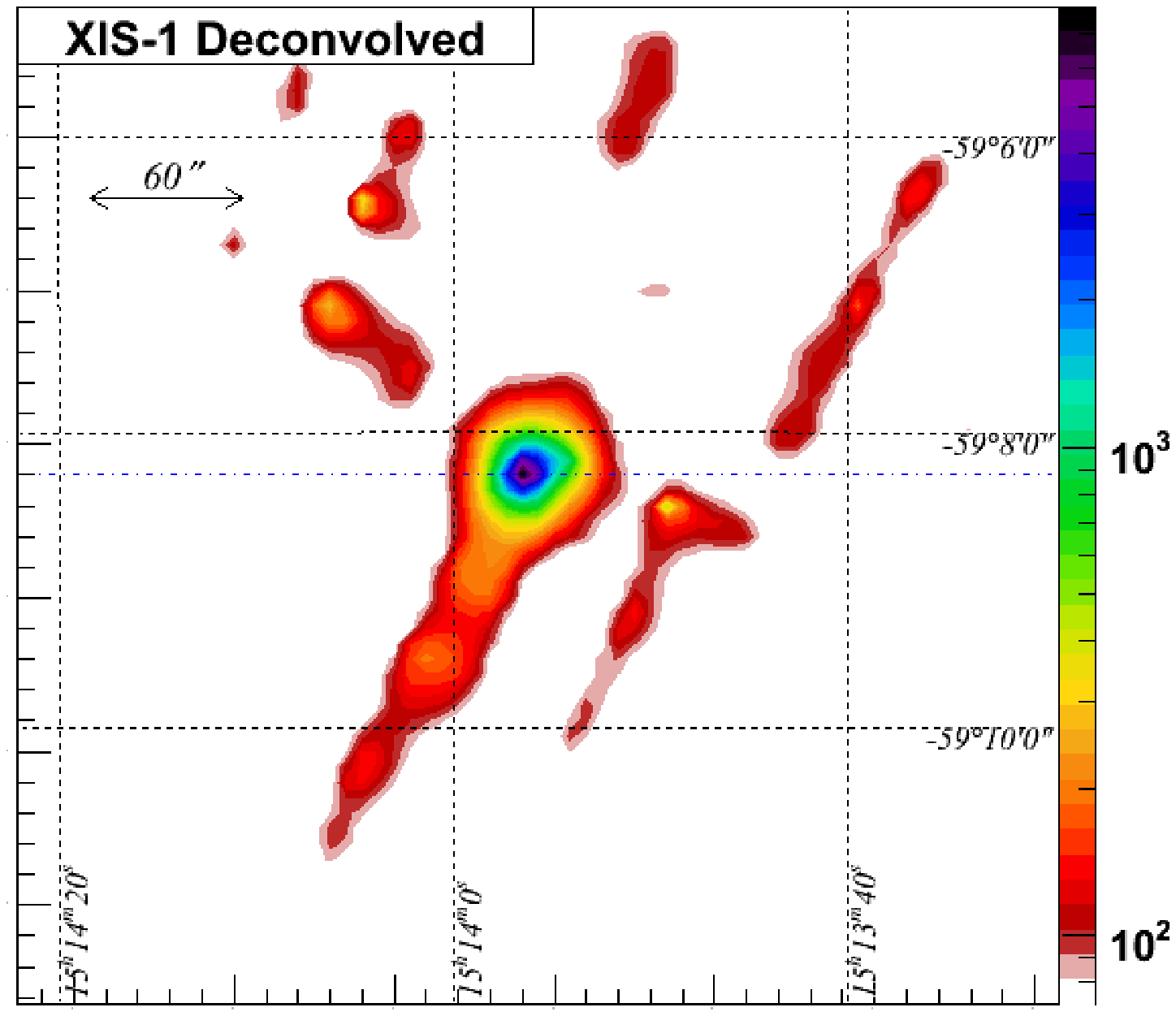}
\FigureFile(5.5cm,){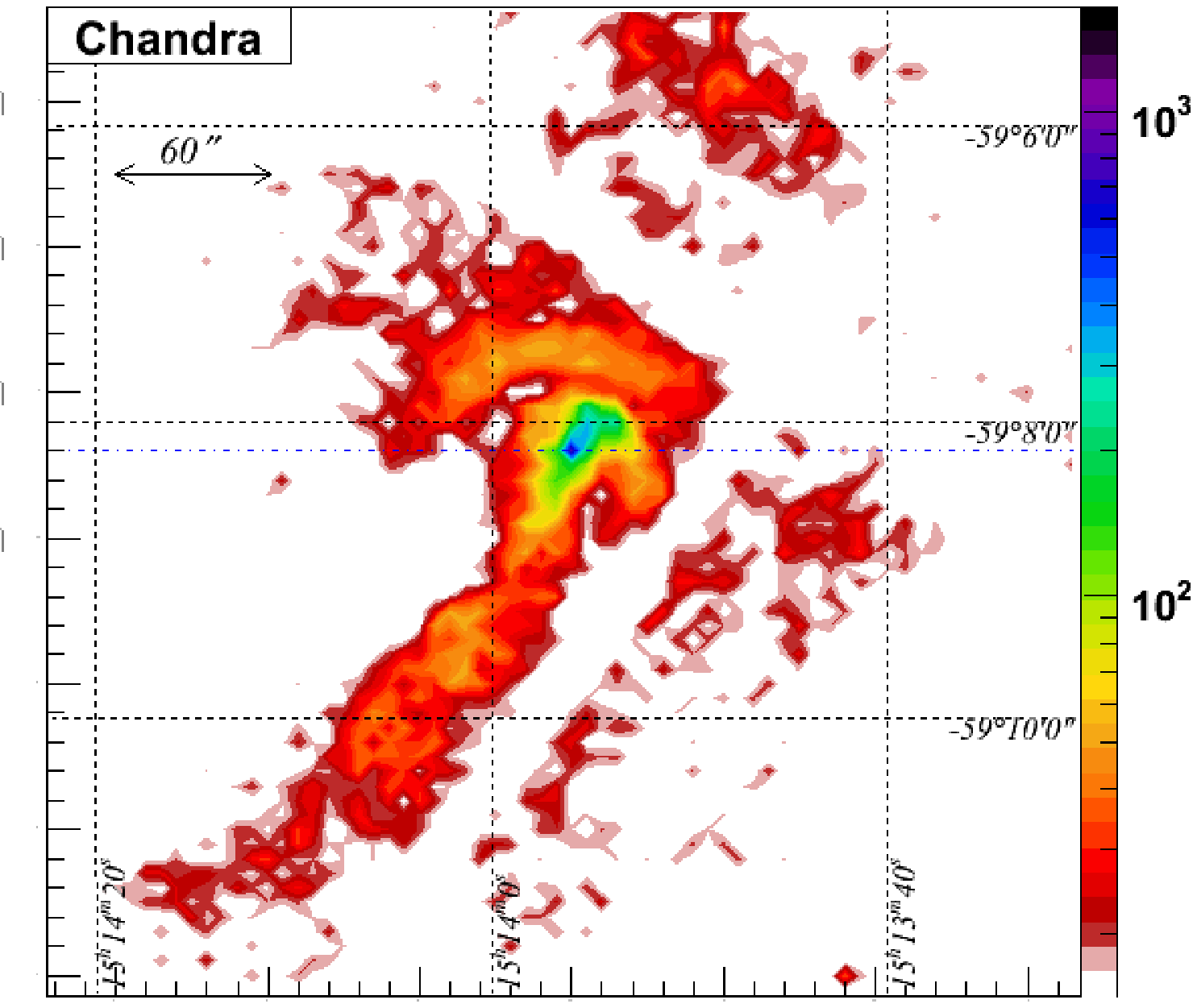}
\FigureFile(5.5cm,){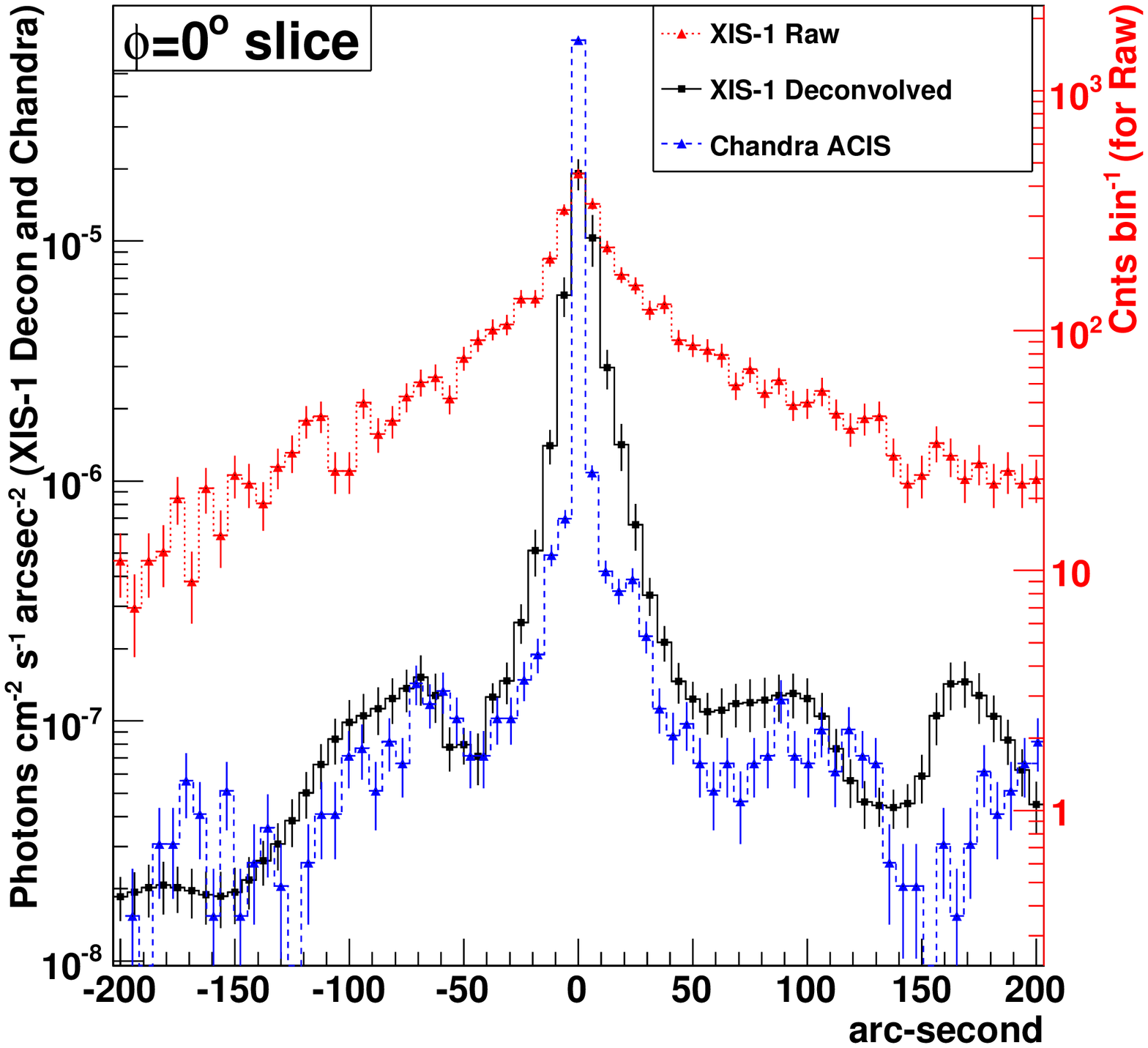}

\begin{center}
\caption{ 
  PSR B1509-58 region: 
  XIS-1 raw image ({\it top left}), 
  convolved Chandra ACIS image with XIS-1 PSF ({\it top center}), 
  XIS-1 0.5-3 keV deconvolved image ({\it bottom left}), 
  Chandra ACIS 0.5-3 keV image ({\it bottom center}),
  and cross-section profiles sliced along Right Ascension
  (blue dash-dot lines in the images) ({\it bottom right}).
  Both  XIS-1 and Chandra ACIS 
  images are binned with a same tile size of $\sim 6''\times 6''$.  
  A pixel of the central core of PSR B1509-58 in the Chandra image 
  is corrected for the pile-up effect.
  Error bars in cross-section profiles represent
  1-$\sigma$ photon-statistics uncertainties.
}

\label{fig:psr1509}

\end{center}
\end{figure}

Figure \ref{fig:psr1509} shows the raw XIS-1 image and 
the deconvolved image in the 0.5--3 keV band,
and the Chandra ACIS image with that convolved with the XIS-1 PSF.
These cross-section profiles along the Right Ascension axis
are also shown.
The surface brightness was calculated assuming a power-law 
spectrum with photon index $\Gamma =1.8$.  The
Chandra image suffers from the pile-up effect at the position of PSR
B1509-58.
The sliced surface brightness profile of the deconvolved image is
consistent with the Chandra image if the pixel saturation is 
corrected using the XIS-1 image. 

The jet-like structure extending from the pulsar is restored in the 
deconvolved image. However other extended features around the pulsar 
are missed or incorrectly restored. Their surface brightness is 
lower than 1/50 of the peak value, or below the dynamic range of 
this deconvolution method. 

\subsubsection{RCW 89: 0.5--3 keV}
\begin{figure}

\FigureFile(5.5cm,){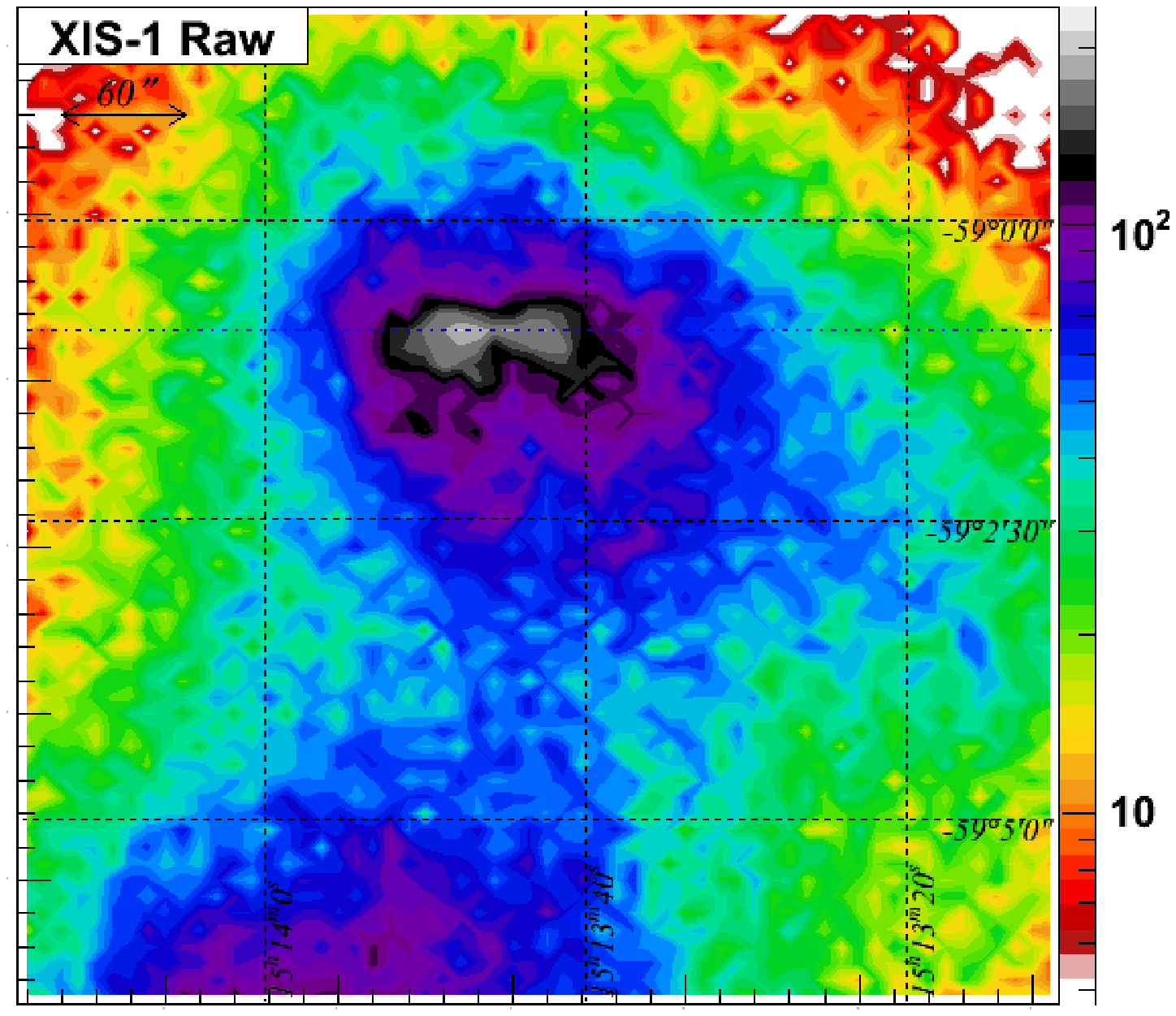}
\FigureFile(5.5cm,){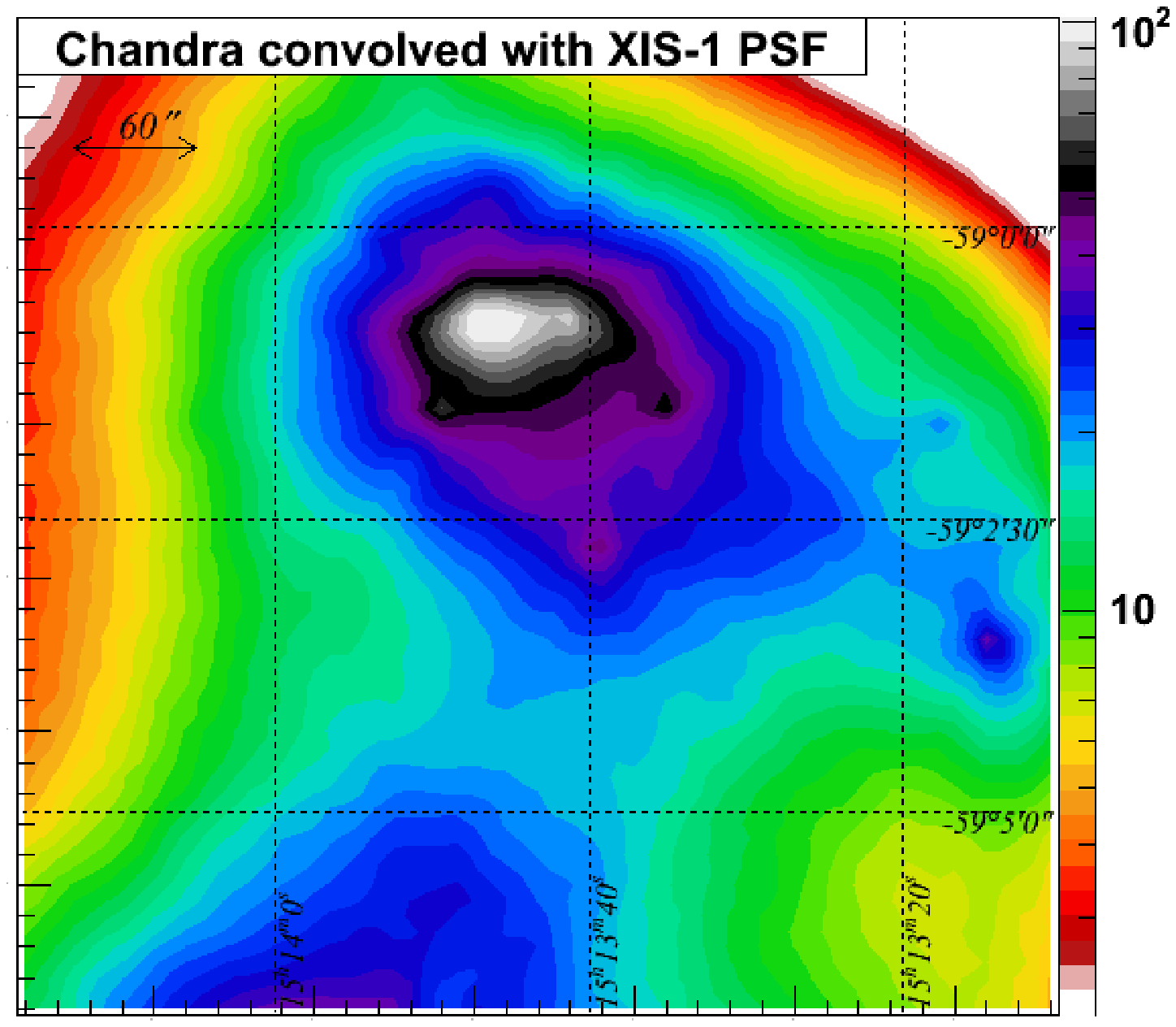}

\FigureFile(5.5cm,){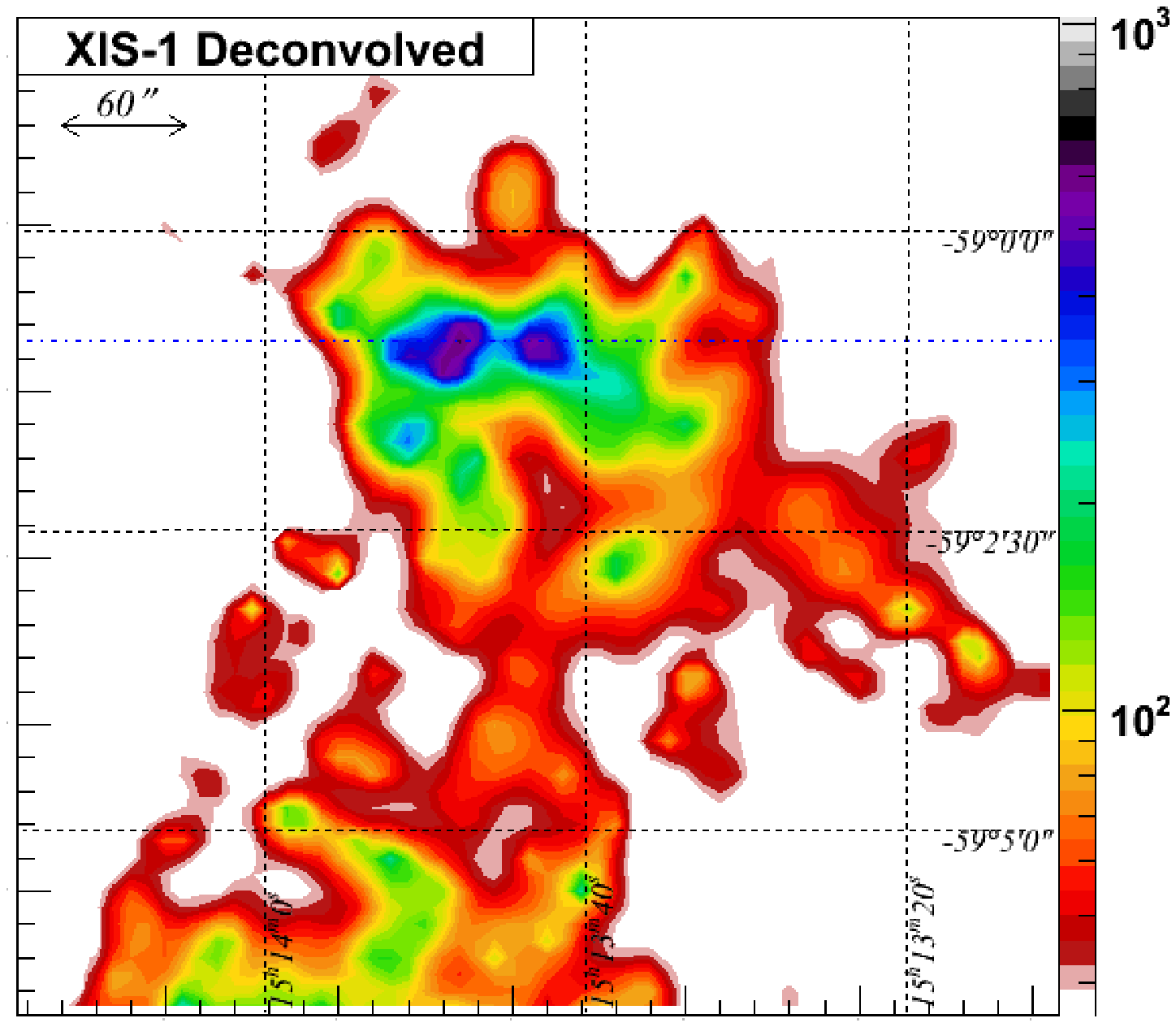}
\FigureFile(5.5cm,){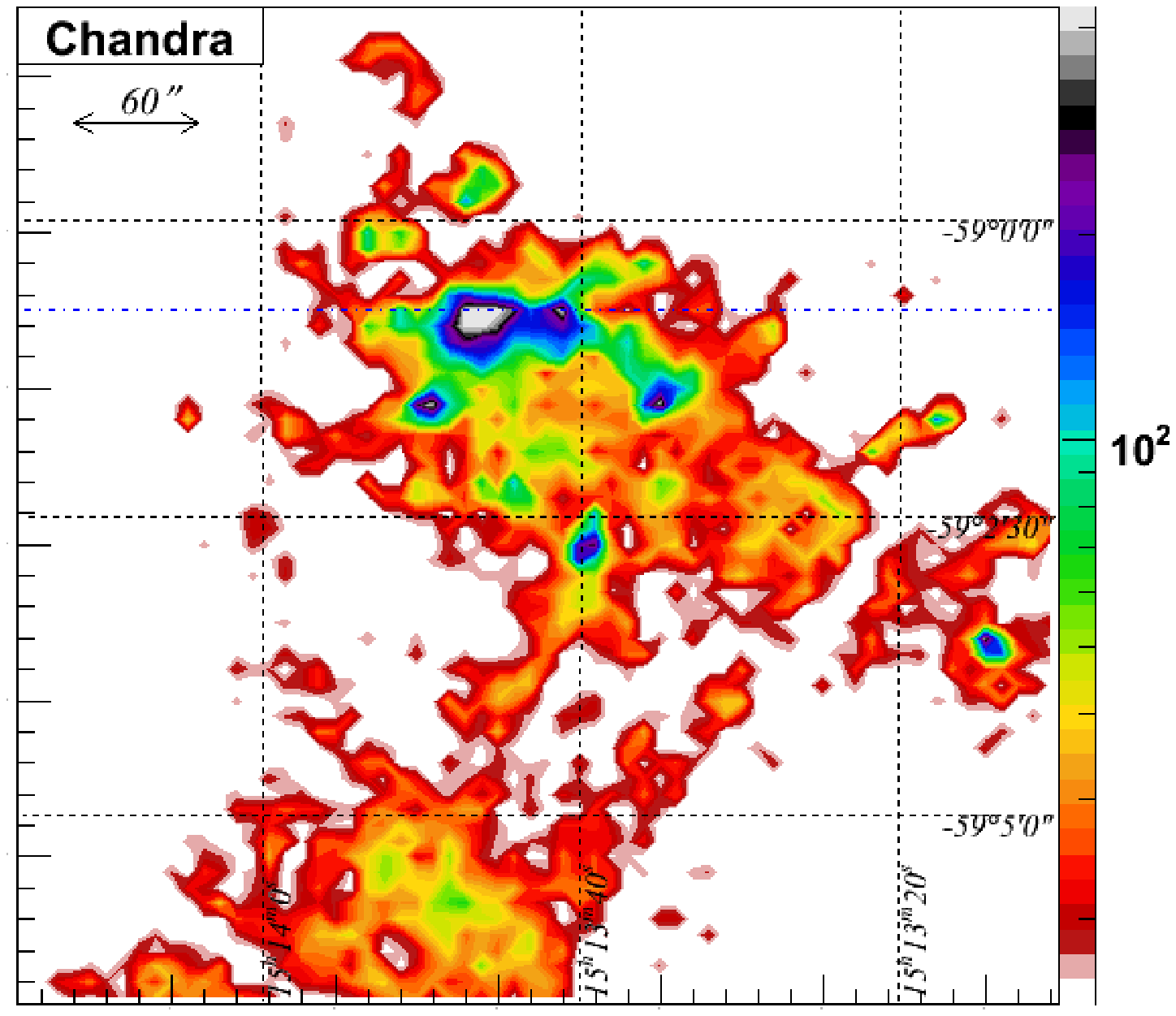}
\FigureFile(5.5cm,){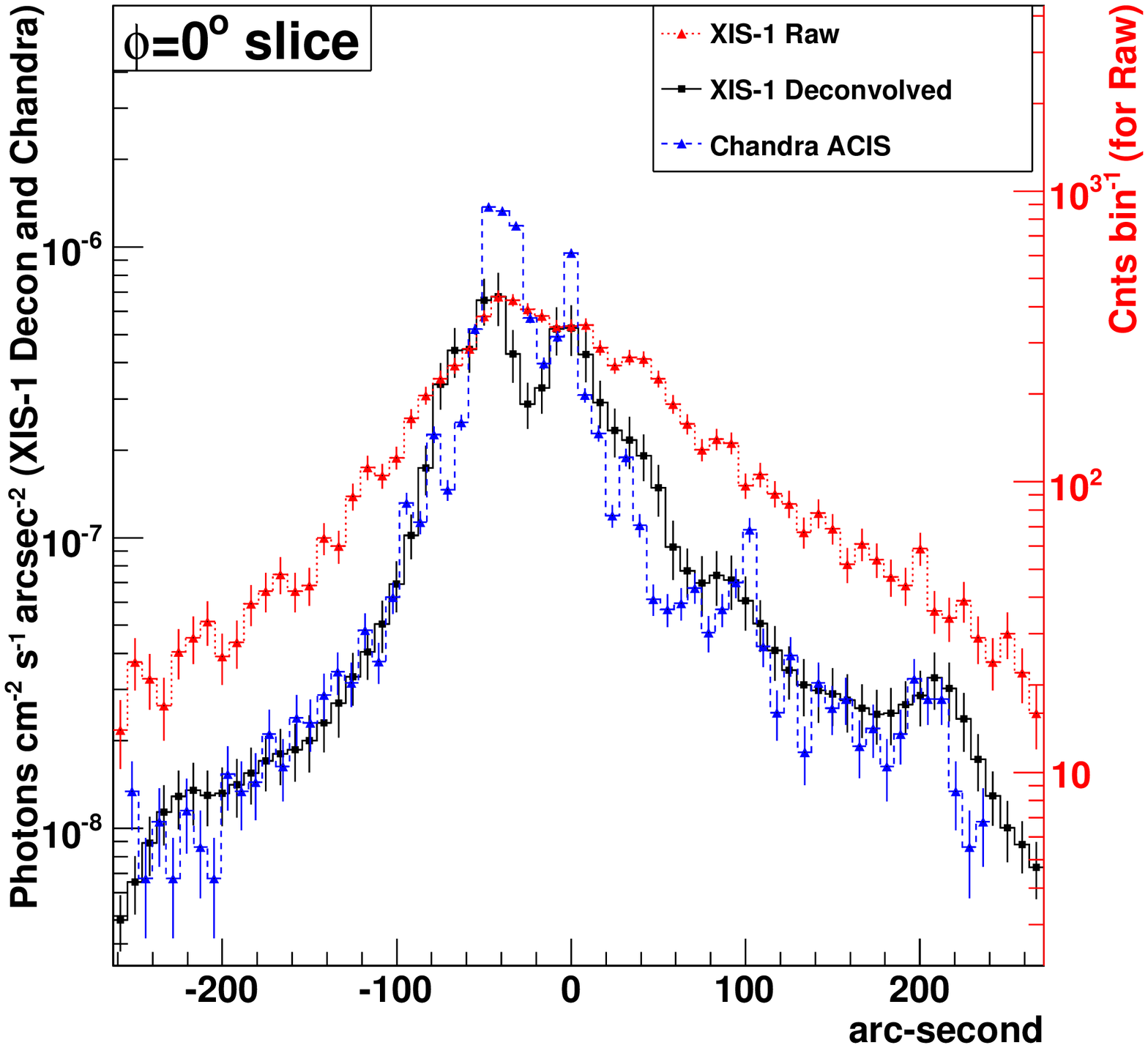}

\begin{center}
\caption{ RCW 89 region: 
  XIS-1 raw image ({\it top left}), 
  convolved Chandra ACIS image with XIS-1 PSF ({\it top center}), 
  XIS-1 0.5-3 keV deconvolved image ({\it bottom left}), 
  Chandra ACIS 0.5-3 keV image ({\it bottom center}),
  and cross-section profiles sliced along Right Ascension
  (blue dash-dot lines in the images) ({\it bottom right}).
  Both XIS-1 and Chandra images
  are binned with a same tile size of $\sim 8''\times 8''$.
  Error bars in peak cross-section profiles represent
  1-$\sigma$ photon-statistics uncertainties.
}
\label{fig:rcw89}
\end{center}
\end{figure}

%
Figure \ref{fig:rcw89} shows the raw XIS-1 image and 
the deconvolved image of RCW 89 in the 0.5--3 keV band,
and the Chandra ACIS image with that convolved with the XIS-1 PSF.
These cross-section profiles along the Right Ascension axis
are also shown.
For this source, $8\times 8$ raw pixels were combined to one tile 
to secure high numbers of photons per tile.
The xissim PSF model was used in the response matrix inversion 
because the image area extended larger than $6'$ 
from the XRT optical axis. 
The surface brightness is calculated assuming a power-law 
spectrum with photon index $\Gamma =1.8$.
The extended bright region of scale greater than $\sim 20''$ 
in the upper part of the image has been reproduced well in the 
deconvolved XIS-1 image but positions of narrower high points 
do not agree with those in the Chandra image. 
We interpret this due both to the inaccuracy of the xissim PSF model 
and to the limited angular resolution of $\sim 20''$. 
Within these limitations, the deconvolution method has reproduced 
the prominent shell-like structure in the upper part. 

\section{Conclusion and Future Prospect}
\label{sec:conc}

We have developed an image deconvolution method for the Suzaku XIS
based on response matrix inversion and adaptive smoothing.  The method
has been tested with two XIS-1 images both containing extended sources
and one prominent point source: Cen A and PSR B1509-58/RCW 89.  
By comparing the deconvolved images with the corresponding Chandra ACIS 
images, we conclude that spatial resolution has been restored to 
$\sim 20''$ to a brightness level around 1/50 of the 
brightest tile in the image. 

Recent X-ray instruments including Suzaku are finding complex
morphology of thermal, non-thermal and K-shell line emissions in many
extended sources including young supernova remnants (SNRs)
(e.g. \cite{Ueno2007}), pulsar wind nebulae (PWNs)
(e.g. \cite{Seward2006}), the Galactic Center region
(e.g. \cite{Koyama2007b,Koyama2007c,Koyama2007d}), and galaxy clusters
(e.g. \cite{Sanders2005}).  To understand such objects, the spatial 
resolution of Suzaku XIS has to be improved substantially.
The present work provides a procedure for such improvement.

We plan to improve PSF modeling and incorporate the XRT alignment scheme 
developed by the Suzaku XIS team.

\vspace{0.5cm}

We are grateful to the Suzaku team and SLAC/KIPAC members
for their support for the present work.  
Special thanks are due to Drs K. Makishima,
K. Mitsuda, Y. Ogasaka, T. Takahashi, R. Blandford, S. Kahn, G. Madejski,
and H. Tajima. We thank the anonymous referee for valuable 
comments. 
This work has been carried out under supports of 
the US Department of Energy contract to SLAC No. 
DE-AC3-76SF00515, Kavli Institute for Particle and Astrophysics
and Cosmology (KIPAC) at Stanford University,
and Japanese Ministry of Education, Culture, Sports, 
Science and Technology (MEXT), Grant-in-Aid No. 18340052.


%
%
%
%
%


\end{document}